\documentstyle[12pt,epsf]{article}

\renewcommand{\theequation}{\thesection.\arabic{equation}}

\begin{document}
\renewcommand{\thefootnote}{\arabic{footnote}}
\setcounter{footnote}{0}
\begin{titlepage}
\renewcommand{\thefootnote}{\fnsymbol{footnote}}
\makebox[2cm]{}\\[-1in]
\begin{flushright}
\begin{tabular}{l}
CERN--TH/95--65/R\\
UM--TH--95--07/R\\
hep-ph/9503492\\
March 1995
\end{tabular}
\end{flushright}
\vskip0.4cm
\begin{center}
{\Large\bf
Resummation of Running Coupling Effects in\\[4pt]
Semileptonic B Meson Decays and Extraction of $|V_{cb}|$
}

\vspace{1.3cm}

Patricia Ball$^1$,
M. Beneke$^2$ and
V.M.\ Braun$^3$\footnote{On leave of absence from St.\ Petersburg
Nuclear Physics Institute, 188350 Gatchina, Russia.}

\vspace{1.3cm}

$^1${\em CERN, Theory Division, CH--1211 Gen\`{e}ve 23, Switzerland}\\[0.4cm]
$^2${\em Randall Laboratory of Physics,
University of Michigan, Ann Arbor, Michigan 48109, USA}\\[0.4cm]
$^3${\em DESY, Notkestra\ss e 85, D--22603 Hamburg, Germany}

\vspace{1.5cm}

{\bf Abstract\\[5pt]}
\parbox[t]{\textwidth}{
We present a determination of $|V_{cb}|$ from semileptonic B
decays that includes resummation of supposedly large perturbative
corrections, originating from the running of the strong coupling.
We argue that the low value of the BLM scale found previously for
inclusive decays is a manifestation of the renormalon divergence of
the perturbative series starting already in third order. A reliable
determination of $|V_{cb}|$ from inclusive decays is possible if one either
uses a short-distance b quark mass or eliminates all unphysical
mass parameters in terms of measured observables, such that all
infra-red contributions of order $1/m_b$ cancel explicitly.
We find that using the $\overline{\rm MS}$ running mass significantly
reduces the perturbative coefficients already in low orders.
%The scheme and scale dependence of the results is reduced.
For a semileptonic branching ratio of $10.9\%$ we obtain
$|V_{cb}|(\tau_B/1.50\,{\rm ps})^{1/2}=
0.041\pm 0.002$ from inclusive decays,
in good agreement with the value extracted
from exclusive decays.
}

\vfill

{\em  submitted to Phys.\ Rev.\ D }
\end{center}
\end{titlepage}
\renewcommand{\thefootnote}{\arabic{footnote}}
\setcounter{footnote}{0}
\newpage

\section{Introduction}
The physics of heavy flavours has experienced a rapid development within the
past few years, driven by new data that aim to test
the Standard Model and to determine its fundamental parameters. In particular,
semileptonic B decays for the moment provide the best possibility to
determine the CKM matrix element $|V_{cb}|$. Two competing strategies,
which both have received considerable attention, are
the determination of $|V_{cb}|$ from the total {\em inclusive}
semileptonic decay rate \cite{inclusive}
and from the
{\em exclusive} $B\to D^* l\bar\nu$ decays at the
point of zero recoil \cite{NThUp}.
In both cases the absence of $1/m_b$ corrections allows
an accurate theoretical description. The decay rates
can be calculated within perturbation theory up to terms of order $1/m_b^2$.
Moreover, the $1/m^2_b$ corrections are estimated to be rather
small ($\sim 5\%$). Thus, at present, the theoretical accuracy of the
determination
of $|V_{cb}|$ is to a large extent limited by a poor control over
perturbative radiative corrections, which are only known to one-loop accuracy.
An explicit calculation of the second order correction is a
very hard enterprise already for $b\rightarrow u$ transitions and
even more so for $b\rightarrow c$ transitions because of the c quark mass,
whose numerical effect is very important, see
\cite{HP84,BBBG94}.

A process of major phenomenological interest is the total inclusive
B meson decay rate with a c quark in the final state, which is calculable
in perturbation theory as
\begin{equation}\label{defGamma}
    \Gamma(B\to X_c e \bar \nu) = \Gamma_0(a)\left[
      1-C_F\frac{\alpha_s}{\pi} g_0(a)+O(\alpha_s^2)\right],
\end{equation}
where $C_F=4/3$ and $a = (m_c/m_b)^2$.

The tree-level decay rate, including the phase space factor $f_1(a)$, reads
\begin{eqnarray}
  \Gamma_0(a)& = &\frac{G_F^2 m_b^5}{192\pi^3} f_1(a)\,,
\nonumber\\
f_1(a) &=& 1-8a+8a^3-a^4-12a^2\ln a,\label{def:f1}
\end{eqnarray}
and the function $g_0(a)$ is known in analytic form \cite{Nir89}.
 Here and below $m_b$ and $m_c$ denote pole masses.
The perturbative expression in
(\ref{defGamma}) should be complemented by  non-perturbative corrections
 suppressed by powers of the heavy quark masses \cite{OPE},
and we will take these corrections into account in the final
analysis. In the major part of the paper, however, we
restrict ourselves to perturbation theory and estimate
higher-order perturbative corrections to (\ref{defGamma}).

\begin{figure}[t]
 \vspace*{-5cm}
 \centerline{\epsffile{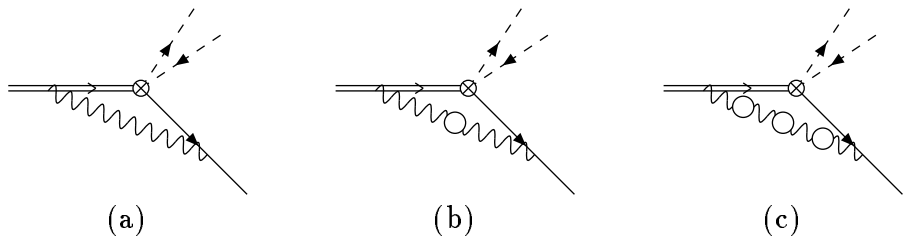}}
 \vspace*{-21.5cm}
\caption[]{Generic radiative corrections for heavy-particle decays: (a)
leading order, (b)~with a fermion bubble insertion, and (c) with a chain of
fermion bubbles. The broken lines represent the lepton pair produced
in the decay.}\label{fig:1}
\end{figure}

For the realistic value $m_c/m_b=0.3$,  Luke, Savage and Wise \cite{LSW94}
have given an estimate for the $\alpha_s^2$ correction in
 (\ref{defGamma}) as
\begin{equation}\label{LSW0.3}
        1-1.67 \frac{\alpha_s(m_b)}{\pi}
        -15.1 \left(\frac{\alpha_s}{\pi}\right)^2
\end{equation}
where  $\alpha_s$ is the
$\overline{\rm MS}$ coupling.
For $m_c/m_b\to 0$, i.e.\ for $b\to u$
decays, the estimated size of the second order correction is even more
striking \cite{LSW94}:
\begin{equation}\label{LSW0.0}
        1-2.41 \frac{\alpha_s(m_b)}{\pi}
-28.7 \left(\frac{\alpha_s}{\pi}\right)^2.
\end{equation}
The coefficients in front of $\alpha_s^2$ were in both cases obtained
by an explicit calculation of the diagrams corresponding
to the insertion
of a fermion loop into the gluon line
in the leading-order virtual correction, as in Fig.~\ref{fig:1}(b),
or the splitting of an emitted gluon into a light quark-antiquark pair, for
the real emission. These contributions
are proportional to the number of light fermion flavours and
the above numerical estimates are obtained by
restoring the full one-loop QCD $\beta$-function by the substitution
$N_f \to N_f-33/2$.
This replacement assumes the hypothesis of BLM \cite{BRO83}
that the dominating radiative corrections originate from the running of
the strong coupling. The result of this procedure is usually expressed
as a redefinition of the scale of the coupling in the leading-order
correction that completely absorbs the second order correction. The
magnitude of these corrections leads to very low BLM
scales for semileptonic decays \cite{LSW94}:
\begin{equation}
   \mu_1^{b\to u} = 0.07m_b, \quad   \mu_1^{b\to c} = 0.13m_b,
\end{equation}
with numerical values of order (350--650)$\,$MeV that are hardly acceptable.
The authors of Ref.~\cite{LSW94} interpreted their result as an indication
that an accurate determination of $|V_{cb}|$ from inclusive decays requires
knowledge of the exact second order correction and even those of higher
order. In Ref.~\cite{NEU94} the large two-loop correction was interpreted as
a breakdown of perturbation theory which disfavours the inclusive
approach to $|V_{cb}|$ in comparison to the exclusive one, for which
large radiative corrections do not appear in the same approximation
\cite{NEU94}. On the other hand, as noted in \cite{SU95}, the
difference in the size of the $\alpha_s^2$ correction for inclusive and
exclusive decays largely
disappears, when the scale of the leading-order correction
is chosen equally as $\sqrt{m_c m_b}$ in both cases. Still, the very fact
of low BLM scales suggests the investigation of yet higher order
radiative corrections.

It is this question we address in this paper. Our analysis extends the
results of Ref.~\cite{LSW94} for inclusive decays and repeats that
of \cite{Nnew} for exclusive ones in that we
resum the effects due to one-loop running of the strong coupling,
but to all orders in perturbation theory. Thus, our investigation
of higher order corrections assumes the dominance of vacuum polarization
effects also in higher orders and we do not address
the question whether knowledge of the exact two- (and higher) loop
corrections as compared to the BLM approximation is important. The
idea is that {\em if} higher order corrections are large {\em at a certain
scale}, they are presumably dominated by running coupling effects and can
thus be taken into account exactly, at least within the restriction to
one-loop running. The remaining corrections are then small
 and therefore can only be accounted for by an exact
calculation. Formally, we resum terms of the type
$\alpha_s(\beta_0\alpha_s)^n$,
of which the correction found in \cite{LSW94,NEU94} is the first term with
$n=1$. These can be traced by a calculation of
contributions proportional to $N_f^n$ given by
a chain of fermion loops as in Fig.~\ref{fig:1}(c).
The leading-order BLM  scales calculated in
\cite{LSW94,NEU94} correspond to
using the QCD coupling at some characteristic
virtuality obtained
by averaging $\ln k^2$, where $k$ is the gluon momentum, over the
leading-order diagram. The resummation that we perform in this paper
amounts to averaging with the
one-loop running coupling
$\alpha_s(k^2)$ itself, rather than $\ln k^2$.
We have developed a technique
to implement this resummation in Refs.~\cite{BB94b,I} and refer
the reader to these articles
for all conceptual and technical issues that we do not repeat in
the present application to semileptonic B decays.

We find that the large second order radiative correction
to $b\to ue\bar\nu$ transitions calculated in
\cite{LSW94} is in fact already close to the regime, where
the series starts to diverge because of
factorially growing coefficients. In our approximation
(called `Naive Non-Abelianization' in \cite{BB94b,I}) the
series in (\ref{LSW0.0}) is continued as
 \begin{equation}\label{LSWcontinue}
        1-2.41 \frac{\alpha_s}{\pi}\Bigg\{1
+11.12\left(\frac{\alpha_s}{\pi}\right)
+149.3 \left(\frac{\alpha_s}{\pi}\right)^2
+2319 \left(\frac{\alpha_s}{\pi}\right)^3
+42751 \left(\frac{\alpha_s}{\pi}\right)^4+\ldots\Bigg\},
\end{equation}
and with $\alpha_s(m_b)=0.21$ one gets a non-convergent series of
corrections to the decay rate already in low orders:
 \begin{eqnarray}\label{result1}
    \Gamma(B\to X_u e \bar \nu) &=& \Gamma_0(0)\left\{
        1-2.41 \frac{\alpha_s(m_b)}{\pi}
\Big[1+0.75+0.67+0.70+0.87+1.27+\ldots\Big]\right\}
\nonumber\\
    &=& \Gamma_0(0)\left\{
        1-2.41 \frac{\alpha_s(m_b)}{\pi}\Big[2.31\pm 0.62\Big]\right\}
 \nonumber\\
    &=& (0.63\pm 0.10)\,\Gamma_0(0).
\end{eqnarray}
In attributing a numerical value to this divergent series we assume
 that it is asymptotic. Then one must truncate it at the
minimal term and  its value gives an estimate of the intrinsic limitation
of the perturbative calculation,\footnote{
In practice, we adopt a similar procedure based on the Borel integral,
and give the principal value of the Borel integral as the central value,
and the imaginary part (divided by $\pi$) as an estimate of
the uncertainty, see Sec.~2
for details.} which cannot be reduced by computing higher orders.

For $b\to ue\bar\nu$ transitions the minimal term occurs at third order in
$\alpha_s$ and its size is comparable to the second order correction.
Numerically, it gives a 15\% uncertainty for the total decay rate,
not taking into account the uncertainty in the input parameters
$\alpha_s$ and the quark masses. For
$b\to ce\bar\nu$ transitions, in the same approximation of vacuum polarization
dominance, we obtain
\begin{equation}
\Gamma(B\to X_c e\bar{\nu}) = (0.77\pm 0.05)\, \Gamma_0(0.3)
\end{equation}
with a 7\% uncertainty. We also note that the uncertainties
associated with the fixed-sign
factorial divergence of perturbative expansions
cannot be reduced by the use of
 a different renormalization scheme, or a change of scale in
the coupling \cite{BEN93}. Thus, the change of scale
suggested in Ref.~\cite{SU95}, which decreases the second order coefficient,
is ineffective already at the next order, because the reduction of
coefficients is compensated by an increase of $\alpha_s$.

However, it would be premature to draw a pessimistic conclusion from
the apparently bad behaviour of perturbative corrections. The large
corrections displayed above originate from infra-red regions in the
integration over loop momenta and produce an uncertainty
parametrically of order
$\Lambda_{\rm QCD}/m_b$. As it turns out, the importance of infra-red
regions is solely due to the choice of an input parameter -- the pole
mass as renormalized mass parameter --, which is incompatible with the
short-distance properties of the decay process. The series that relates
the pole to the bare mass contains large finite
renormalizations of infra-red origin. If these are made explicit -- for
example using the $\overline{\rm MS}$ renormalized mass -- they cancel
with the large corrections of infra-red origin present in the
perturbative series for the decay width \cite{BIG94,BBZ94}.

The preference of the
$\overline{\rm MS}$ (or another `short-distance')
mass might seem surprising and even
counter-intuitive. After all, it is the pole mass that governs the
(partonic) decay kinematics and it is the visualization of an almost
on-shell (up to effects of order $\Lambda_{\rm QCD}$) b quark inside the
meson that motivated the approximation of the meson
decay by a free quark decay in the first place.
But this picture also implies the
existence of a static field (in the rest frame of the quark) around the
heavy quark, which behaves as $1/r$ at short distances. Thus a
contribution of order $\Lambda_{\rm QCD}$ to the self-energy of the quark is
stored at large distances, $r\sim 1/\Lambda_{\rm QCD}$, of the order of the
radius of the heavy-light meson. However, due to the
Kinoshita-Lee-Nauenberg cancellations,
in an inclusive decay of a heavy quark the decay vertex is localized
to within a distance $1/m_b$ and the energy stored in the field at
large distances $r\ge 1/\mu$ (where $m_b > \mu \ge \Lambda_{\rm QCD}$ is
a factorization scale) cannot participate in the hard process, but is
absorbed into a rearrangement of the colour field of the hadronizing
spectator quark. This explains, loosely spoken, why a short-distance
mass is more appropriate in the description of inclusive decays as a
hard process. This reasoning assumes
that the quark produced by the weak current is fast and does not apply
to a massive quark produced with zero recoil (cf. Sec.~3.2).

Since ultimately any quark mass parameter is unphysical, the most
transparent way to exhibit the infra-red cancellations  would be to
eliminate any mass parameter in terms of a suitable physical quantity,
provided it is determined by short distances and does not import
$1/m_b$ corrections (which rules out meson masses). However, as with
the strong coupling constant, it is convenient to use a mass parameter
for book-keeping purposes and as a {\em numerical\/} input parameter.
It is only at this point that the $\overline{\rm MS}$ mass
(or any other mass
defined at short distances) is favoured over the pole mass, because
it does not import infra-red effects.

In fact, the divergence of the series of radiative corrections
to the decay rates is only one aspect of the problem with using the pole mass.
Another aspect is that it cannot be accurately extracted from
measurable quantities through perturbative expansions \cite{BIG94,BB94}.

To reinforce this point we imagine that we used  pole masses as
numerical input parameters, determined from another measurement.
Then one would always find that the size of perturbative corrections
does not allow a determination of the masses to an accuracy better
than $\pm 100\,$MeV (we quote the estimate from \cite{I}).
Treating the uncertainty in the input parameters as uncorrelated
with the uncertainty in the theoretical prediction for the radiative
corrections to the B decay width, we obtain a $\pm$21\% and $\pm$10\%
uncertainty for $B\to X_u e\bar{\nu}$ and $B\to X_c e\bar{\nu}$ decays,
respectively. This translates into an irreducible theoretical
uncertainty of $\pm$10\% in $|V_{ub}|$ and $\pm$5\% in $|V_{cb}|$.
The $10\%$ uncertainty in $\Gamma(B\to X_c e\bar{\nu})$ is to be compared
with the present experimental uncertainty of 12\% of the branching ratio.

The calculation of  decay rates with running masses
has already been considered in \cite{BN94}
at the level of one-loop perturbative corrections.
After resummation, we find that apart
from the cancellation of infra-red contributions the
perturbative coefficients are strongly reduced already in low orders.
In particular, if the $\overline{\rm MS}$ running mass is used in the
tree-level decay rate, the series of radiative corrections for
$b\to ue\bar\nu$ decays becomes
 \begin{equation}\label{LSWcontinueMS}
        1+4.25 \frac{\alpha_s}{\pi}\Bigg\{1
+8.99\left(\frac{\alpha_s}{\pi}\right)
+35.15 \left(\frac{\alpha_s}{\pi}\right)^2
+241.1 \left(\frac{\alpha_s}{\pi}\right)^3
+1547 \left(\frac{\alpha_s}{\pi}\right)^4+\ldots\Bigg\}.
\end{equation}
Although the leading-order correction has increased (and changed sign),
the higher-order coefficients are significantly reduced, so that
with the same value of $\alpha_s$ as above we get the well convergent
series
 \begin{eqnarray}\label{result2}
    \Gamma(B\!\to \!X_u e \bar \nu) &=& \overline{\Gamma}_0(0)\left\{\!
        1+4.25 \frac{\alpha_s(m_b)}{\pi}
\Big[1+0.604+0.159+0.073+0.032+0.022+\ldots\Big]\!\right\}
\nonumber\\
    &=& \Gamma_0(0)\left\{
        1+4.25 \frac{\alpha_s(m_b)}{\pi}\Big[1.92\pm 0.01\Big]\right\}
\end{eqnarray}
with an uncertainty that is negligible compared to the uncertainty
inherent in the restriction to vacuum polarization corrections.

It should be noted that anticipating the eventual cancellation of
infra-red regions we can define the numerical value of the sum of
radiative corrections even when it diverges, with some (ad hoc)
prescription. Provided we
use the same prescription to define the pole mass, and provided we know that
the cancellation occurs, we can then simply delete nearly all large
uncertainties. Thus, the calculation in the on-shell scheme
(using pole masses) can
be saved at the cost of introducing
consistent {\em non-perturbative\/} prescriptions to define the pole mass and
to sum the series of radiative corrections. We shall consider this
(conceptually less appealing) possibility
as well, and shall see that after resummation of
$\beta_0^n\alpha_s^{n+1}$ corrections the decay rates calculated
in the on-shell and $\overline{\rm MS}$ schemes are close to each other,
 provided
the same short-distance mass is used as an input parameter. This
supports {\em a posteriori} the assumption that the
dominant higher order corrections are taken into account by vacuum
polarization effects.

We conclude that the large corrections found in \cite{LSW94} do not endanger
the accuracy of the theoretical treatment of inclusive decays. We do find,
however, large corrections beyond second order in $\alpha_s$, in particular
in the on-shell scheme, defined in the sense of the previous paragraph.
We suggest that these corrections are more important than the
$\alpha_s^2$ corrections left out by the restriction to the effects
of running coupling. Although the accuracy of this restriction
is not known and is certainly the main deficiency of our analysis,
we believe that resummation of one-loop running effects provides a fair
estimate of higher order perturbative corrections and the corresponding
`M-factors' (defined below)
should be incorporated into any phenomenological analysis.

As already emphasized above,
 an appropriate treatment of quark masses as input
parameters is of equal importance as the size of radiative corrections
to the width itself. Most previous determinations of $|V_{cb}|$ from
inclusive decays have used the OS scheme and pole masses as
numerical input. Instead,  we
choose the $\overline{\rm MS}$ masses as numerical input parameters.
Thus, when we use the OS scheme in the above sense, the pole masses
are calculated from the $\overline{\rm MS}$ masses. This procedure
is not without its own difficulties, since we must rely on direct
determinations of $\overline{\rm MS}$ masses, as from QCD
sum rules, which have been obtained without the resummation which we
implement for the decay width.

We use our results for a new
determination of $|V_{cb}|$ from both inclusive
and exclusive decays. We find a good agreement between the
determination in the $\overline{\rm MS}$ and OS scheme, and obtain
 as our final value
\begin{equation}
    |V_{cb}|(\tau_B/1.50\,{\rm ps})^{1/2}=0.041\pm 0.002,
\end{equation}
where the combined error comes from several sources that will be detailed
below.

The presentation is organized as follows.
In Sec.~2 we review the necessary formulas for resummation.
Section~3 contains our main results for the
BLM-improved perturbative series in B meson semileptonic decays.
These two sections are more technical and those readers interested only
in results may continue with Sec.~4 directly.
The updated
analysis of $|V_{cb}|$ obtained with the resummed formula along with its
major uncertainties is given in Sec.~4, while
Sec.~5 is reserved for a summary and conclusions. Some technical
discussion and especially long formulas are given in the appendices.
As a new analytic result, we derive
an expression for the total $b\to u$ semileptonic decay width with a non-zero
gluon mass, which provides the input necessary for resummation of
running coupling effects \cite{BB94b,I}.

\section{General formulas}
\setcounter{equation}{0}

We now formulate the resummation in precise terms. We are interested in the
 effect of radiative corrections to the total semileptonic width, which
we define as
\begin{equation}
\Gamma(B\to X_c e \bar \nu)=\Gamma_0\Bigg\{1
 - C_F\frac{\alpha_s(m_b)}{\pi}g_0(a) \Big[1 +
\sum_{n=1}^\infty \tilde d_n(a)\alpha_s^n(m_b)\Big]\Bigg\}\,.
\end{equation}
The functions $\tilde d_n(a)$ depend on the ratio of the charm and bottom
quark masses, and  are polynomials in the number of light
flavours $N_f$:
\begin{equation}\label{Nf-expand}
       \tilde d_n(a) = \tilde d_{n0} +\tilde d_{n1}N_f +\ldots+
\tilde d_{nn} N_f^n\,.
\end{equation}
The coefficient $\tilde d_{nn}(a)$ comes from the insertion of $n$ fermion
loops in the gluon lines in the leading-order correction;
this is the quantity we
calculate explicitly.
Substituting $N_f\to N_f-33/2$ in the highest power of $N_f$,
we rewrite (\ref{Nf-expand}) as
\begin{equation}\label{collect}
       \tilde d_n(a) = \delta_n(a) + (-\beta_0)^n d_n(a)\,,
\end{equation}
where the (uncalculated) $\delta_n$  is  by construction at most of order
$N_f^{n-1}$. We use the definition
of the first coefficient
$\beta_0$ of the QCD $\beta$-function including the factor $-1/(4\pi)$,
\begin{equation}
       \beta_0 = -\frac{1}{4\pi}\Big(11-\frac{2}{3}N_f\Big)
\end{equation}
with $N_f=4$ for the case of interest\footnote{We neglect the charm
quark mass in quark loops. Its effect is small \cite{I}.}.
The standard BLM prescription \cite{BRO83} uses
$d_1$ to fix the scale in the coupling in the leading-order
correction. In the generalization developed in \cite{BB94b,Nnew,I} all
terms $\delta_n$ in (\ref{collect}) are neglected, but all $d_n$ are kept
as an estimate of radiative corrections for arbitrary $n$.
Then we get
\begin{equation}
\Gamma(B\to X_c e \bar \nu)=\Gamma_0\Bigg\{1
   -C_F\frac{\alpha_s(m_b)}{\pi}g_0(a)\Big[1+
\sum_{n=1}^\infty (-\beta_0)^n d_n(a)\alpha_s^n(m_b)\Big]\Bigg\}\,.
\end{equation}
Note that because of the factor $1/(4\pi)$ in $\beta_0$ the
expansion parameter is effectively $\alpha_s/(4\pi)$.
To quantify the effect of partial summation of $N$ orders, we
introduce the `M-factors'
\begin{eqnarray}\label{def-M}
           M_N^{b\to c}[a,-\beta_0\alpha_s(m_b)] &\equiv&
   1+\sum_{n=1}^N (-\beta_0)^n d_n(a)\alpha_s^n(m_b)\,,
\nonumber\\
 M_\infty^{b\to c}[a,-\beta_0\alpha_s(m_b)] &\equiv&
M_{N\to\infty}^{b\to c}[a,-\beta_0\alpha_s(m_b)]\,,
\end{eqnarray}
that measure the modification of the leading-order radiative correction
by integrating with the running coupling at the vertex.
The limit $N\to\infty$ in  Eq.~(\ref{def-M}) does not exist
in a rigorous sense, reflecting the
factorial divergence of the coefficients $d_n$ in high orders.
Assuming that the perturbative series is asymptotic, one is led to the
conclusion that the uncertainty in the summation is in fact power-suppressed
in $m_b$, and can be estimated numerically.
Thus, in the following, numerical values of $M_{\infty}$ will always be given
with an uncertainty, reflecting this problem. This uncertainty cannot
be eliminated without a rigorous factorization of the corresponding
infra-red contributions into the matrix elements of higher dimensional
operators.

An equivalent way to present the results is to absorb the M-factors
into a redefinition of the scale in the lowest-order correction\footnote{
For finite $N$  one must
expand (\ref{def-BLM}) in $\alpha_s$ and truncate
the expansion at the desired order.}
\begin{eqnarray}\label{def-BLM}
   \alpha_s(\mu_N^{b\to c}) &\equiv& \alpha_s(m_b)
M_N^{b\to c}[a,-\beta_0\alpha_s(m_b)]\,,
\nonumber\\
       \mu_\infty^{b\to c} &=& \mu_{N\to\infty}^{b\to c}.
\end{eqnarray}
The scale $\mu_1^{b\to c}$ is just the leading-order
  BLM scale studied in \cite{LSW94} and the $\mu_n$ with $n>1$
correspond to a more accurate treatment
of the distribution in the gluon virtuality, reflected by the size
of higher-order corrections with up to $n$ fermion loops.
The uncertainty in the summation of the series is translated to the
uncertainty in the ultimate BLM scale $\mu_\infty$ \cite{BB94b}.

\begin{figure}[t]
\vspace{-5cm}
\centerline{\epsffile{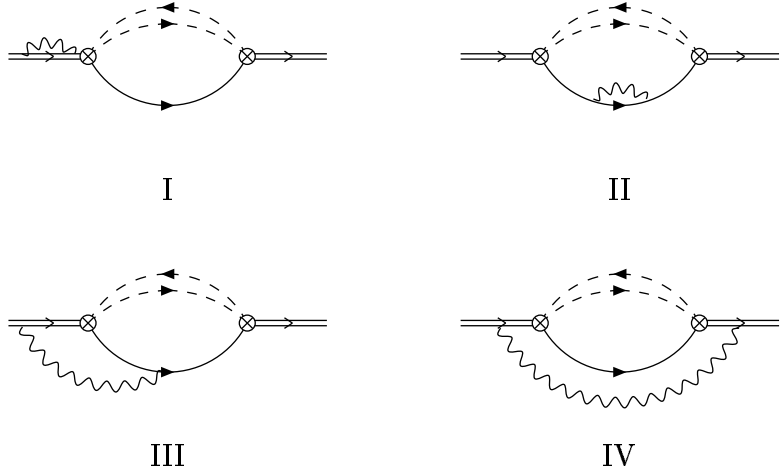}}
\vspace*{-18.4cm}
\caption[]{Leading-order radiative corrections to the
transition operator, whose imaginary part gives the inclusive semileptonic
decay width. Double line: b quark, solid line: c quark, dashed lines:
leptons.}\label{fig:2}
\end{figure}

The calculation of
the coefficents $d_n(a)$ requires the evaluation
of diagrams such as those in Fig.~\ref{fig:1}, with the
insertion of $n$ fermion loops in the
gluon lines. This problem is solved in a most economical way by  applying
a dispersion technique, and reduces to the calculation
of the leading-order diagrams with finite gluon mass $\lambda$.
Denote by
$$-\Gamma_0(a)C_F \alpha_s/\pi\,g_0(a)d_0(a,\lambda^2)$$
 the sum of the diagrams
in Fig.~\ref{fig:2} calculated
with a finite gluon mass $\lambda$, so that
$d_0(a,\lambda^2=0)=1$.
For the contribution from  the one fermion loop insertion, Smith and Voloshin
\cite{SV94} have
derived a useful representation (in the V scheme of \cite{BRO83})
\begin{equation}\label{Smith}
  d^V_1(a) = -\int_0^\infty \frac{d\lambda^2}{\lambda^2}
\left(d_0(a,\lambda^2)-\frac{m_b^2}{\lambda^2+m_b^2}\right)\,,
\end{equation}
which has been used in the analysis of Ref.~\cite{LSW94}.
The calculation of diagrams with multiple
fermion loop insertions involves
precisely the same function $d_0(a,\lambda^2)$,
and can thus be done at little additional calculational
expense. In particular, the fixed-order coefficients $d_n(a)$ are
obtained as \cite{BB94b,I}
\begin{eqnarray}\label{final1}
    d_n(a) &=&  \frac{d^n}{du^n} B[D](a,u)_{|_{u=0}}\,,
\nonumber\\
B[D](a,u) &=&
 -\frac{\sin(\pi u)}{\pi }\int_0^\infty \frac{d\lambda^2}{\lambda^2}\,
\left(\frac{\lambda^2}{\mu^2}e^C\right)^{-u}
 \left[d_0(a,\lambda^2)-1\right]
\nonumber\\&=&
 -\frac{\sin(\pi u)}{\pi u }\int_0^\infty d\lambda^2\,
\left(\frac{\lambda^2}{\mu^2}e^C\right)^{-u}
 d'_0(a,\lambda^2),
\end{eqnarray}
where $ d'_0(a,\lambda^2) =(d/d \lambda^2)d_0(a,\lambda^2)$
and $C$ is a scheme-dependent finite renormalization constant.
In the $\overline{\rm MS}$ scheme one has $C=-5/3$,
in the V scheme  $C=0$.
It is easy to check that for $n=1$ the above expression reproduces
Eq.~(\ref{Smith}).

A closed expression can be derived for the sum of all
diagrams with an arbitrary number of fermion bubble insertions
\cite{BB94b,I}:
\begin{equation}\label{rBS}
(-\beta_0\alpha_s) M_\infty[a,-\beta_0\alpha_s] =
\int_0^\infty d\lambda^2\, \Phi(\lambda^2)\,d'_0(a,\lambda^2)
+[d_0(a,\lambda_L^2)-1],
\end{equation}
where $\alpha_s =\alpha_s(\mu)$,
\begin{equation}\label{Phi}
\Phi(\lambda^2) = {}-\frac{1}{\pi}\arctan\left[\frac{-\beta_0\alpha_s\pi}
{1-\beta_0\alpha_s\ln(\lambda^2/\mu^2 e^C)}\right] -
\,\theta(-\lambda_L^2-\lambda^2)\, ,
\end{equation}
and
\begin{equation}\label{Landau}
\lambda_L^2 =- \mu^2\exp[1/(\beta_0\alpha_s)-C]
\end{equation}
is the position of the Landau pole in the strong coupling.
Note that the term with the $\theta$ function exactly cancels the
jump of the $\arctan$ at $\lambda^2=-\lambda_L^2$, so that $\Phi(\lambda^2)$
is a continuous function of $\lambda^2$.

In this paper we cannot give a detailed discussion of the assumptions
underlying  the derivation of Eqs.~(\ref{final1})
and (\ref{rBS}), and refer the  reader to the
corresponding sections in \cite{BB94b,I}. Still, two short comments are
appropriate.

First, note that the product
$\alpha_s(\mu) M_\infty[a,-\beta_0\alpha_s(\mu)]$ is explicitly
scale-invariant, provided the running of the coupling is implemented to
leading logarithmic accuracy: $\alpha_s(\mu_1)=\alpha_s(\mu_2)/
(1-\beta_0 \alpha_s(\mu_2)\ln(\mu_1^2/\mu_2^2))$. This result is also
scheme-invariant, and in particular  independent of the
renormalization constant $C$, provided the couplings are
consistently related in the same BLM approximation,
that is by keeping only the terms with highest power in $N_f$.
This is in contrast to the finite order summation coefficients $M_N$, which
are scheme- and scale-dependent.
In the following we assume the $\overline{\rm MS}$ scheme for
the coupling $\alpha_s$, and the
normalization point $\mu=m_b$.

Secondly, notice that the second term in (\ref{rBS}) involves the
radiative correction to the decay rate with a finite gluon mass, analytically
continued to the Landau pole $\lambda^2_L<0$. The renormalon divergence
of the perturbation theory is reflected \cite{BBZ94} by non-analytic
terms in the expansion of $d_0(a,\lambda^2)$ at small $\lambda^2$ and
leads to an imaginary part in this continuation.
The size of the imaginary part (divided by $\pi$), $\delta M_\infty
\equiv 1/(\pi|\beta_0|\alpha_s)\,\mbox{Im}\,d_0(a,\lambda^2_L)$, yields
an estimate of the ultimate accuracy of perturbation theory, beyond
which it has to be complemented by non-perturbative corrections.
The real part of (\ref{rBS}) coincides with the
sum of the perturbative series defined by the principal value of the Borel
integral \cite{I},
and the imaginary part of $d_0(a,\lambda^2_L)$ coincides with the
imaginary part of the Borel integral.

The calculation of the
diagrams in Fig.~\ref{fig:2} with a finite gluon mass is
straightforward, albeit tedious, and has been undertaken in
\cite{LSW94}. Since no formulas were given there,
 we had to redo this calculation. For $b\to ue\bar\nu$ decays, that is
for a massless quark in the final state, we have succeeded in obtaining
an analytic expression for the decay rate. For $b\to ce\bar\nu$ decays,
we leave the answer in form of at most two-dimensional integrals and
evaluate them numerically. The corresponding
formulas are collected in Appendix A.

\section{Resummation of BLM-type radiative corrections}
\setcounter{equation}{0}

Our results are summarized in Table~\ref{tab:massive}.
\begin{table}
$$
\begin{array}{l|rrrrrrrrr}
\sqrt{a} & 0\phantom{.00} & 0.1\phantom{0} & 0.2\phantom{0} & 0.3\phantom{0}
& 0.4\phantom{0} & 0.5\phantom{0} & 0.6\phantom{0} & 0.7\phantom{0}
& 0.99\\ \hline\hline
g_0 & 1.81 & 1.63 & 1.42 & 1.25 & 1.12 & 1.01 & 0.92 & 0.85 & 0.75\\ \hline
d_0 & 1\phantom{.00} & 1\phantom{.00} & 1\phantom{.00} & 1\phantom{.00} &
1\phantom{.00} & 1\phantom{.00} & 1\phantom{.00} & 1\phantom{.00} &
1\phantom{.00}\\
d_1 & 5.34 & 5.00 & 4.55 & 4.09 & 3.59 & 3.10 & 2.62 & 2.16 & 1.18\\
d_2 & 34.4 & 30.9 & 27.3 & 23.9 & 20.5 & 17.4 & 14.3 & 11.4 & 5.24\\
\hline
M_0 & 1\phantom{.00} & 1\phantom{.00} & 1\phantom{.00} & 1\phantom{.00}
& 1\phantom{.00} & 1\phantom{.00} & 1\phantom{.00} & 1\phantom{.00}
& 1\phantom{.00}\\
M_1 & 1.75 & 1.70 & 1.64 & 1.57 & 1.50 & 1.43 & 1.37 & 1.30 & 1.17 \\
M_2 & 2.42 & 2.31 & 2.17 & 2.04 & 1.90 & 1.77 & 1.65 & 1.53 & 1.27 \\
M_\infty & 2.31 & 2.35 & 2.24 & 2.10 &
1.96 & 1.82 & 1.69 & 1.57 & 1.38\\
         & \ \pm 0.62 &\ \pm 0.52 &\ \pm 0.44 &\ \pm 0.37 &
\ \pm 0.32 & \ \pm 0.26 & \ \pm 0.21 & \ \pm 0.15 & \ \pm 0.03\\
\hline
     \mu_1/m_b & 0.07 & 0.08 & 0.10 & 0.13 & 0.17 & 0.21 & 0.27 & 0.34
       & 0.55\\
\mu_\infty/m_b & 0.13 & 0.13 & 0.14 & 0.15 & 0.17 & 0.20 & 0.23 & 0.27
       & 0.37\\
\hline\hline
\end{array}
$$
\caption[]{Resummation of $\beta_0^n\alpha_s^{n+1}$ corrections for
semileptonic B decay widths, see text.
The values of $M_n$ and $\mu_n$ are given for
$-\beta_0^{(4)}\alpha_s(m_b)=0.14$.
}\label{tab:massive}
\end{table}
For several representative values of $\sqrt{a}\equiv m_c/m_b$ we give
calculated values of the fixed-order coefficients $d_n$, the
partial sums $M_n$ and the BLM scales $\mu_n$. For $n=1$,
the $\beta_0\alpha_s^2$ correction, our results coincide with
the ones obtained in \cite{LSW94}. We now discuss
the numbers in Table~\ref{tab:massive} in detail.

\subsection{Hierarchy of BLM scales}
The BLM scale $\mu_\infty$ can be
larger than the leading-order scale $\mu_1$. This may come unexpectedly.

The (leading-order) BLM prescription uses the average of $\ln k^2$, the
average virtuality of the gluon,
 as the scale in the coupling. In high orders,
this substitution generates a series of radiative corrections with
a geometric growth of coefficients $d_n \sim (d_1)^n$ to be
compared with the factorial growth of the exact coefficients.
Resummation to all orders corrects for this discrepancy by adjusting
$\mu_\infty$ so that the expansion of $\alpha_s(\mu_\infty)$ gives
the correct $d_n$ for all $n$. Since for this reason for large $n$
the true $d_n$ will always outgrow $ (d_1)^n$ and since the
series is with fixed sign, one might suspect that the usual BLM
scale-setting rather underestimates higher-order corrections.

For small c quark masses the effect is
opposite, and
the very small leading BLM
scale $\mu_1$ is simply an artefact of
truncating the perturbative expansion at
low order.
The scales $\mu_1$ and
 $\mu_\infty$ are given by
\begin{eqnarray}\label{compare}
   \mu_1 &=&
m_b \,\exp[-d_1/2] =
m_b\,\exp\left[\frac{1}{2\beta_0\alpha_s}(M_1-1)\right]\,,
\nonumber\\
   \mu_\infty &=& m_b\,\exp\left[\frac{1}{2\beta_0\alpha_s}
   \left(1-\frac{1}{M_\infty}\right)\right]\,.
\end{eqnarray}
Although $\delta M \equiv M_\infty -M_1$ is positive in all cases
we consider,
the expression in parentheses in the first equation in (\ref{compare})
is numerically larger than the similar expression in the
second equation, provided $\delta M < (M_1-1)^2/(2-M_1)$. This is
satisfied for small mass ratios, see Table~\ref{tab:massive}, and results in
 $\mu_\infty
>\mu_1$ (recall that $\beta_0$ is negative with our definition).

Note that the scale  $\mu_\infty$
is bounded from below: as long as  $M_\infty$ is positive, $\mu_\infty$
is larger than $m_b \exp[1/(\beta_0\alpha_s(m_b))]$,
the position of the
infra-red Landau pole in the running coupling.
There is no such restriction for $\mu_1$, which can
take values below the pole. Again, this is an artefact of the
truncation at fixed order.

\subsection{Suppression of infra-red contributions for finite c quark
mass}
With a massive quark in the final state
the radiative corrections apparently are reduced: The M-factors
$M_n$ are smaller, and the BLM scales $\mu_n$ are larger.
This
has been observed in \cite{LSW94} for the BLM scale in leading order, and
it continues to all orders, although the difference between
$m_c/m_b=0$ (relevant to $b\to ue\bar\nu$ transitions) and $m_c/m_b=0.3$
is less pronounced
after resummation.
The ambiguities related to the summation of a divergent
series are also reduced and almost vanish in the limit of zero
recoil $m_c\to m_b$. One can understand this by continuing the argument
given in the introduction. As explained, these ambiguities arise,
because the use of the pole mass parameter implies a static picture,
while the energy stored in the field at large distances cannot
be converted into hard radiation in the weak decay process. This
assumed that the produced quark is fast in the rest frame of the
initial b quark. When $m_c\to m_b$, the c quark is slow. Then, since the
long-range part of the field of the b quark is universal, it
can be smoothly transferred
to the c quark and is simply irrelevant for the description of the
decay. Therefore these long-distance contributions cannot be seen in
the form of ambiguities in the zero-velocity limit.

\begin{figure}[t]
\epsfxsize=0.5\textwidth
\centerline{\epsffile{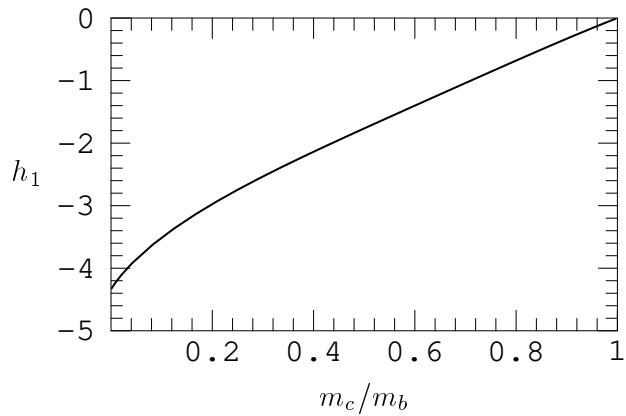}}
\caption[]{Coulombic contributions to the B decay rate,
  Eq.~(\protect\ref{h1}), expressed in terms
of pole masses, as a function of the mass ratio $m_c/m_b$.}\label{fig:3}
\end{figure}

To see this more explicitly,
we recall that contributions of small momenta to decay rates can be traced
by non-analytic terms in the expansion at small values of the gluon mass
\cite{BBZ94}. For the leading-order radiative correction to the B decay
width this expansion takes the form\footnote{Note that terms $O(\lambda^2\ln
\lambda^2)$ are absent \cite{BBZ94}. This is explained by the absence of a
renormalon ambiguity in the kinetic energy of the heavy quark inside
the B meson, at least in the approximation considered here.}

\begin{equation}
   d_0(a,\lambda^2) = 1+h_1(a)\sqrt{\frac{\lambda^2}{m_b^2}}+
                      h_2(a)\frac{\lambda^2}{m_b^2} +
\Big[ h_{31}(a)\ln (\lambda^2/m_b^2) +h_{32}(a)\Big]
\left(\frac{\lambda^2}{m_b^2}\right)^{3/2}
                      + O(\lambda^4 \ln\lambda^2)
\end{equation}
with
\begin{eqnarray}\label{h1}
   h_1(a) &=& {}-\frac{\pi}{2 f_1(a)g_0(a)}
            \Big[5-16a^{1/2}-24a-24 a^{3/2}
            +24a^2+48a^{5/2}-8a^3-8a^{7/2}+3a^4
\nonumber\\&&{}
            -48a^{3/2}\ln a -12 a^2\ln a \Big]\,.
\end{eqnarray}
Note that the tree-level phase space factor $f_1(a)$ and the
leading-order radiative correction $g_0(a)$ are extracted.
The function $h_1(a)$ is plotted as a function of the mass ratio
$\sqrt{a}=m_c/m_b$ in Fig.~\ref{fig:3}.
%\phantom{\ref{fig:1}}
For the realistic value $m_c/m_b=0.3$
it is reduced by approximately a factor~2 compared to the massless case.

These long-range contributions to the static field
are responsible for the
major part of the ambiguity in the sum of the perturbative series.
Within our approach, this ambiguity is related to the imaginary part of
$d_0(a,\lambda^2)$, continued analytically to the position of the Landau
pole (\ref{rBS}), (\ref{Landau}), and equals
\begin{equation}\label{errorbar}
 (-\beta_0\alpha_s)\delta M_\infty =h_1(a)\frac{\lambda_L}{m_b}
   +O(\lambda_L^3/m_b^3).
\end{equation}
The decrease of the value for $M_\infty$ at $\sqrt{a}=0.3$
($b\to ce\bar\nu$ decays) compared to $a=0$ ($b\to ue\bar\nu $ decays)
roughly equals the decrease of the uncertainty.

In fact, these infra-red contributions are spurious, and can be removed
by re-expressing the decay widths in terms of the short-distance (say,
$\overline{\rm MS}$)
b and c quark masses instead of the pole masses
 \cite{BIG94,BBZ94}. To trace this
cancellation we write e.g.\ the b quark pole mass $m_b$ as related to the
running $\overline{\rm MS}$ mass
$\overline{m}_b$ by the perturbative expansion
\begin{equation}\label{mass-lambda}
   m_b = \overline{m}_b(\overline{m}_b)
   \Bigg\{1+C_F\frac{\alpha_s(\overline{m}_b)}{\pi}
 r_0(\lambda^2)+\ldots
\Bigg\}\,,
\end{equation}
where we keep a finite gluon mass $\lambda$ as for the decay width.
The expansion of
$r_0(\lambda^2)$ at small gluon masses reads \cite{BIG94,BBZ94}:
 \begin{equation}\label{mass-small}
              r_0(\lambda^2)=1-\frac{\pi\lambda}{2m_b}+\ldots
\end{equation}
Using Eqs.~(\ref{mass-lambda}), (\ref{mass-small})
and similar expressions for the c quark, we find that,
when calculated with a small
gluon mass,
the tree level decay rate is modified to
\begin{equation}
     \Gamma_0(a,\lambda) =\Gamma_0(a)\left\{1-C_F\frac{\alpha_s}{\pi}
       \frac{\pi\lambda}{m_b}\left[
        5/2+(\sqrt{a}-a)\frac{d}{da}\ln f_1(a)\right]+\ldots\right\}.
\end{equation}
It is easy to see that
the correction linear in $\lambda$ exactly cancels with a similar term
in the radiative  correction to the decay width,
\begin{eqnarray}\label{Gamma-lambda}
    \Gamma &=& \Gamma_0(a,\lambda)
\left[ 1-C_F\frac{\alpha_s}{\pi} g_0(a)d_0(a,\lambda^2)+\ldots\right]
\nonumber\\
 &=&\Gamma_0(a) \left[1-C_F\frac{\alpha_s}{\pi}
\frac{\pi\lambda}{m_b}\left\{
5/2+(\sqrt{a}-a)\frac{d}{da}\ln f_1(a) + \frac{1}{\pi} g_0(a) h_1(a)
\right\}+\ldots\right],
\end{eqnarray}
with the  function $h_1(a)$ defined as above. The terms in curly brackets
add to zero,
so that the total decay rate is free from infra-red contributions to this
accuracy \cite{BIG94,BBZ94}. It is only the use of a pole mass
as an input parameter which introduces
infra-red $1/m_b$ effects in the tree-level
decay rate and in the radiative corrections, which cancel in the
product.
For a massless quark in the final state
several terms in the small-$\lambda$ expansion can
easily be obtained analytically, with the
result
\begin{equation}\label{thirdpower}
    \Gamma =\Gamma_0(0)\Bigg\{1+C_F\frac{\alpha_s}{\pi}
\Bigg[\frac{65}{8}-\frac{\pi^2}{2}
   - \left(\frac{27}{4}+\frac{4\pi^2}{3}\right)\frac{\lambda^2}{m_b^2}
   - \left(\frac{13\pi}{8}+4\pi\ln(\lambda^2/m_b^2)\right)
\frac{\lambda^3}{m_b^3}+\ldots\Bigg]\Bigg\}\,.
\end{equation}
The infra-red contributions now start at order $O(\lambda^3/m_b^3
\ln\lambda^2)$ and are
numerically neg\-ligible\footnote{The remaining (small) uncertainty is
related to contributions of dimension 6 operators to the decay rate,
which produce non-perturbative corrections of order $1/m^3$. They
can be relevant for D decays, see \cite{CHER95,BDS94}. The terms
proportional to $\lambda^3$
in (\ref{thirdpower})  produce an uncertainty of order 10\% in
the decay rate $D\to X e\bar\nu$, which can be taken as an indication
of the minimal size of $1/m_c^3$ corrections.}.
 If so, it is natural to formulate the perturbative calculation
in terms of a mass parameter
defined at short distances, so that large infra-red contributions
 do not appear. We address this task now.

\begin{table}[bt]
\addtolength{\arraycolsep}{2pt}
\renewcommand{\arraystretch}{1.2}
$$
\begin{array}{l|cc|cc|cc}
n & r_n & M_n^{b} &d_n(0)& M_n^{b\to u}
&\overline{d}_n(0)& \overline{M}_n^{b\to u}\\ \hline\hline
0 & 1          & 1     & 1         & 1     & 1       & 1    \\
1 & 4.6861511  & 1.656 & 5.3381702 & 1.747 & 4.3163 & 1.604\\
2 & 17.622650  & 2.001 & 34.409913 & 2.422 & 8.0992 & 1.763\\
3 & 109.85885  & 2.303 & 256.48081 & 3.126 & 26.680 & 1.836\\
4 & 873.92393  & 2.638 & 2269.4131 & 3.997 & 82.262 & 1.868\\
5 & 8839.6860  & 3.114 & 23679.005 & 5.271 & 421.33 & 1.890\\
6 & 105814.28  & 3.911 & 289417.40 & 7.450 & 1656.1 & 1.903\\
7 & 1484968.4  & 5.476 & 4081180.2 & 11.75 & 12135 & 1.916\\
8 & 23740736.  & 8.978 & 65496131. & 21.42 & 52862 & 1.924\\ \hline
\infty & -  & 2.066\pm 0.231 & - & 2.314\pm 0.615 &-& 1.925\pm 0.012\\
\hline\hline
\end{array}
$$
\addtolength{\arraycolsep}{-2pt}
\renewcommand{\arraystretch}{1}
\caption[]{Effect of the elimination of the b quark pole mass on the
radiative corrections to $b\to ue\bar\nu$ decays, see text. The
given values of $M_n$ correspond to $-\beta_0^{(4)}\alpha_s(m_b)=0.14$.
}\label{tab:eliminate}
\end{table}
\subsection{Elimination of the pole mass: $b\to ue\bar\nu$ decays.}
The b quark pole mass is related to the $\overline{\rm MS}$ mass by the
 perturbative
series
\begin{equation}\label{mass-series}
   m_b = \overline{m}_b(\overline{m}_b)
   \Bigg\{1+C_F\frac{\alpha_s(\overline{m}_b)}{\pi}
   \Big[ 1+\sum_{n=1}^\infty \tilde r_n\alpha_s^n(\overline{m})
    \Big]\Bigg\}\,.
\end{equation}
As above, we approximate
\begin{equation}
              \tilde r_n=(-\beta_0)^n  r_n\,,
\end{equation}
where $r_n$ corresponds to contributions of $n$ fermion loops
to the leading-order diagram for the fermion self-energy and can be
calculated using a representation similar to (\ref{final1}), (\ref{rBS})
in terms of the leading-order diagram with a finite gluon mass
$r_0(\lambda^2)$, see \cite{BB94b,I} for details.
The partial sums for the perturbative series truncated at order $N$
are defined as
\begin{eqnarray}\label{def-Mpole}
           M_N^{b}[-\beta_0\alpha_s(m_b)] &\equiv&
   1+\sum_{n=1}^N (-\beta_0)^n r_n\alpha_s^n(m_b)\,,
\nonumber\\
 M_\infty^{b}[-\beta_0\alpha_s(m_b)] &=&
M_{N\to\infty}^{b}[-\beta_0\alpha_s(m_b)]\,.
\end{eqnarray}
The coefficients $r_n$ were calculated in Ref.~\cite{BB94b} and are
given together with the partial sums $M_N^b$
in the second and third columns in Table~\ref{tab:eliminate}.
The perturbative series defining the pole mass is divergent \cite{BIG94,BB94},
which is reflected by the uncertainty in the factors $M_\infty^b$.
The crucial point is that these uncertainties in defining resummed
pole masses are correlated with uncertainties in the resummed
radiative correction to the decay rate, and cancel
against each other, when the pole mass is {\em defined} by its relation
to the short-distance mass as in (\ref{mass-series}) or eliminated
 in favour of the $\overline{\rm MS}$
mass \cite{BIG94,BBZ94}.

In what follows we shall consider both possibilities.
The first one, which we refer to as calculation in the on-shell scheme (OS),
is to define the resummed inclusive decay rate as
\begin{equation}\label{scheme1}
  \Gamma^{OS}(B\to X_u e \bar \nu) =\Gamma_0^{OS}
\left[1-C_F\frac{\alpha_s}{\pi}g_0(0)M_\infty^{b\to u}\right]\,,
\end{equation}
where it is understood that the b quark pole mass appearing in the
tree-level decay rate is substituted by
\begin{equation}\label{scheme2}
   m_b = \overline{m}_b(\overline{m}_b)
   \left[1+C_F\frac{\alpha_s}{\pi}M_\infty^b\right]
\end{equation}
with the factors $M_\infty^{b\to u}$ and $M_\infty^{b}$ given
in Table~\ref{tab:eliminate} and  the uncertainties deleted.

The second possibility is to use the $\overline{\rm MS}$ scheme
from the very beginning.
Using (\ref{mass-series}) we can write
the decay rate as
\begin{equation}\label{b->u}
    \Gamma(B\to X_u e \bar \nu) = \overline{\Gamma}_0(0)\left\{
        1+C_F\frac{\alpha_s(m_b)}{\pi}\overline{g}_0(0)
\left[1+\sum_{n=1}^N(-\beta_0)^n \overline{d}_n(0)\alpha_s^n(m_b)\right]
\right\}\,,
\end{equation}
where
\begin{equation}
   \overline{g}_0(0)= 5-g_0(0)\,,\qquad
\overline{d}_n(0)=-\frac{g_0(0) d_n(0)-5 r_n}{\overline{g}_0(0)} \,,
\end{equation}
and the tree-level decay rate $ \overline{\Gamma}_0$ is expressed
in terms of $\overline{m}_b(\overline{m}_b)$.
Note that the leading-order radiative correction changes sign and
becomes somewhat larger.

The coefficients $\overline{d}_n$ and partial sums
$\overline{M}_N^{b\to u}$ defined
in an obvious way in analogy to (\ref{def-M}) are given in
Table~\ref{tab:eliminate} in comparison to $d_n$ and $M_N^{b\to u}$,
respectively.
It is seen that the coefficients are drastically reduced, and the series
has become well convergent. The remaining infra-red effects, relevant to
the divergence of the perturbative series, are
suppressed by three powers of the b quark mass \cite{BBZ94} and have
become tiny. Most importantly, this improvement is effective already
at $n=2$. We conclude that
perturbative coefficients in the $\overline{\rm MS}$ scheme are likely to be
much smaller than in the OS scheme without restriction to vacuum
polarization corrections, too.

In the framework of a purely perturbative calculation
the use of the OS scheme can only be justified up to the order where
perturbative series diverge.
In all-order resummations such as the one considered in this paper, one
must make sure that the {\em prescription\/} defining the pole mass
in terms of a short-distance mass or any physical quantity
is consistent with the prescription to sum the perturbative
series of radiative corrections to the decay width.
Even in this case, the OS scheme
is somewhat unnatural since it involves large cancellations between
radiative corrections to decay rates and to the pole masses already in
low orders.

The resummed decay rate in the $\overline{\rm MS}$ scheme is
readily obtained  by inserting
(\ref{scheme2}) into (\ref{scheme1}) and expanding up to
 $O(\alpha_s)$:
\begin{equation}\label{scheme3}
  \overline{\Gamma} =\overline{\Gamma}_0
\left\{1+C_F\frac{\alpha_s}{\pi}
\Big[5 M_\infty^b -g_0(0)M_\infty^{b\to u}\Big]\right\}\,.
\end{equation}
The difference between (\ref{scheme1}) and (\ref{scheme3}) is an effect
of order $(C_F\alpha_s/\pi)^2$, which is beyond our accuracy. It is a pure
scheme dependence, resulting from our incomplete perturbative calculation.
Numerically, the difference amounts to about 6\%, which is significantly
smaller  than the $\pm 15\%$ uncertainty for the radiative corrections
noted in the introduction.

\subsection{Elimination of the pole mass: $b\to ce\bar\nu$ decays.}
Expressing the $b\to ce\bar\nu$ decay rate in terms of the running masses is
slightly more cumbersome.
As above, we start with the resummed decay rate in the OS scheme
\begin{equation}\label{gammaOS}
  \Gamma^{OS}(B\to X_c e \bar \nu) =\Gamma_0^{OS}
\left[1-C_F\frac{\alpha_s}{\pi}g_0(a)M_\infty^{b\to c}\right]\,,
\end{equation}
where now
 both the c and b quark masses have to be expressed in terms of the
running masses. Here we want to be somewhat more general, and
introduce running masses at arbitrary scale $\mu$  as
\begin{equation}\label{M(mu)}
  m_{c,b}=\overline{m}_{c,b}(\mu)\Big[
  1+C_F\frac{\alpha_s(\mu)}{\pi}\overline{M}_\infty^{c,b}(\mu)\Big]\,.
\end{equation}
The factors $\overline{M}_\infty^c(\mu)$ and
$\overline{M}_\infty^b(\mu)$ can most easily be
calculated by observing that the pole mass on the left-hand side of
(\ref{M(mu)}) is scheme-invariant, and using the renormalization-group
expression for the running mass
\begin{equation}\label{eq:RGmass}
\bar m(\mu)=\bar m(\bar m)\,\exp\left(\,\,-\frac{1}{\beta_0}
\int_{\alpha_s(m)}^{\alpha_s(\mu)}
\frac{d\alpha}{\alpha^2}\, \gamma_m(\alpha)
\right)\,,
\end{equation}
where to our accuracy we have to expand the exponential to first order, but
keep the mass anomalous dimension to all orders in $\beta_0^n\alpha_s^{n+1}$,
which is known from \cite{PAL84,I} (see Eq.~(4.19) in \cite{I}):
\begin{equation}
 \gamma_m(\alpha_s)=\frac{\alpha_s}{\pi}\Bigg[1-\frac{5}{6}(\beta_0\alpha_s)
 -\frac{35}{36}(\beta_0\alpha_s)^2+\ldots\Bigg].
\end{equation}
Thus, we obtain, e.g.\ for the c quark
\begin{equation}
C_F\frac{\alpha_s(\mu)}{\pi}\overline{M}_\infty^{c}(\mu) =
C_F\frac{\alpha_s(m_c)}{\pi}M_\infty^{c}
+\frac{1}{\beta_0}
\int_{\alpha_s(m_c)}^{\alpha_s(\mu)}
\frac{d\alpha}{\alpha^2}\, \gamma_m(\alpha),
\end{equation}
where $M_\infty^c$ is a factor relating the c quark pole mass to
the running mass $\overline{m}_c(\overline{m}_c)$, defined as in
(\ref{scheme2}).

Collecting everything, we find
\begin{eqnarray}\label{gammaMS}
  \overline{\Gamma}(B\to X_c e \bar \nu)&=&\overline{\Gamma}_0(\bar a,\mu)
\left\{1+C_F\frac{\alpha_s(\mu)}{\pi}
\Big[5 \overline{M}_\infty^b(\mu) -g_0(\bar a)M_\infty^{b\to c}\right.
\nonumber\\&&\left.{}\hspace{1cm}
+2\Big(\overline{M}_\infty^c(\mu)-\overline{M}_\infty^b(\mu)\Big)\bar a
\frac{d}{d\bar a}\ln f_1(\bar a)
\Big]\right\},
\end{eqnarray}
where $\bar a =(\bar m_c(\mu)/\bar m_b(\mu))^2$.
It is worth while to note that the factor that appears
in front of the logarithmic
derivative of the phase space function $f_1$ is scale-independent:
\begin{equation}
\alpha_s(\mu)\Big(\overline{M}_\infty^c(\mu)-\overline{M}_\infty^b(\mu)\Big)=
\alpha_s(m_c)M_\infty^c-\alpha_s(m_b) M_\infty^b
+\frac{\pi}{\beta_0 C_F}
\int_{\alpha_s(m_c)}^{\alpha_s(m_b)}
\frac{d\alpha}{\alpha^2}\, \gamma_m(\alpha)\,.
\end{equation}
The expression in (\ref{gammaMS})
 is to be compared with the leading-order decay rate \cite{BN94}:
\begin{eqnarray}\label{gammaMSLO}
 \overline{\Gamma}^{LO}(B\to X_c e \bar \nu)&=&
 \overline{\Gamma}_0(\bar a,\mu)
\left\{1+C_F\frac{\alpha_s(\mu)}{\pi}\bar g_0(\bar a,\mu)
\right\}\,,\nonumber\\
 \bar g_0(\bar a,\mu) &=& 5-g_0(\bar a)-\frac{15}{4}\ln\frac{m_b^2}{\mu^2}
-\frac{3}{2}\bar a \ln \bar a \frac{d}{d\bar a}\ln f_1(\bar a)\,.
\end{eqnarray}

To quantify the effect of resummation, we introduce the
corresponding M-factor by rewriting the resummed result in (\ref{gammaMS})
as
\begin{eqnarray}\label{Mbc-bar}
  \overline{\Gamma}(B\to X_c e \bar \nu)
& =  & \overline{\Gamma}_0(\bar a,\mu)
\left\{1+C_F\frac{\alpha_s(\mu)}{\pi}\bar g_0(\bar a,\mu)
\overline{M}^{b\to c}_{\infty}(\mu)\right\}\,.
\end{eqnarray}
We give the corresponding values in Table~\ref{tab:inputMS} for
$\mu=\overline{m}_b$. Tracing the scale dependence of
$\overline{M}_\infty(\mu)$ is
misleading in this case because
$ \bar g_0(\bar a,\mu)$ is strongly scale-dependent.

\subsection{The SV limit and exclusive B decays}

The inclusive decay rate in the
Shifman-Voloshin (SV) limit $m_b,\, m_c\gg m_b-m_c\gg \Lambda_{\rm\scriptsize
QCD}$ \cite{SVlimit} is dominated by  two exclusive
decay channels, $B\to D e \bar\nu$ and $B\to D^* e \bar\nu$. Since the final
state meson is produced almost at rest
(in leading order in the heavy quark expansion), the techniques of the
Heavy Quark Effective Theory are applicable, yielding the decay rate
\begin{equation}
\Gamma(B\to D e \bar\nu,D^* e \bar\nu ) = \frac{G_F^2 |V_{cb}|^2
(m_b-m_c)^5}{60\pi^3}\left(\eta_V^2 +3\eta_A^2\right),
\end{equation}
where $\eta_V$ and $\eta_A$ are the short-distance matching coefficients
of the QCD heavy-heavy currents to the corresponding currents in the
effective theory (at zero recoil):
\begin{equation}
\bar c \gamma_\mu b = \eta_V \bar h_c\gamma_\mu h_b + O(1/m^2),\quad
\bar c \gamma_\mu\gamma_5 b = \eta_A \bar h_c\gamma_\mu\gamma_5 h_b +
O(1/m^2).
\end{equation}
They are given by a perturbative series, in which, as above, we only
keep the BLM-type $\beta_0^n\alpha_s^{n+1}$ terms ($z=m_c/m_b$):
\begin{equation}\label{eq:3.39}
\eta_{V,A} = 1 + C_F\,\frac{\alpha_s(m_b)}{\pi}\,r_0^{V,A}(z)
\left[1+\sum_{n=1}^\infty (-\beta_0)^n d_n^{V,A} \alpha_s^n(m_b)\right]
\end{equation}
with \cite{politics,SVlimit}
\begin{equation}
      r_0^V(z) = -\frac{3}{2}-\frac{3}{4}\,\frac{1+z}{1-z}\,\ln\,z,
\quad
      r_0^A(z) = -2-\frac{3}{4}\,\frac{1+z}{1-z}\,\ln\,z\,.
\end{equation}
The resummation of $\beta_0^n\alpha_s^{n+1}$ terms for the $\eta$'s
was discussed in some detail in \cite{NEU94,Nnew} and can equally
easily be implemented within our dispersion technique.
For completeness, we collect the necessary formulas in Appendix~B.
The results are summarized in Table~\ref{tab:exkl}, where we give
the leading-order coefficients $d_1^{V,A}$ and the resummed enhancement
factors
\begin{equation}
   M^{V,A}_\infty[a,-\beta_0\alpha_s(m_b)]\equiv
1+\sum_{n=1}^\infty (-\beta_0)^n d_n \alpha_s^n(m_b)\,.
\end{equation}
\begin{table}
\renewcommand{\arraystretch}{1.2}
\addtolength{\arraycolsep}{3pt}
$$
\begin{array}{c|cc|cc}
m_c/m_b & d_1^V & d_1^A & M_\infty^V & M_\infty^A\\ \hline\hline
0.1 & 2.47 &  1.63 & 2.07 \pm 0.05 & 2.69 \pm 1.09\\
0.2 & 1.78 &  3.32 & 1.50 \pm 0.07 & 2.37 \pm 0.42\\
0.3 & 1.36 &  2.56 & 1.30 \pm 0.06 & 1.96 \pm 0.16\\
0.4 & 1.08 &  2.17 & 1.20 \pm 0.05 & 1.76 \pm 0.09\\
0.5 & 0.86 &  1.90 & 1.14 \pm 0.04 & 1.64 \pm 0.06\\
0.6 & 0.68 &  1.70 & 1.09 \pm 0.03 & 1.56 \pm 0.05\\
0.7 & 0.52 &  1.53 & 1.06 \pm 0.03 & 1.50 \pm 0.04\\
0.8 & 0.39 &  1.39 & 1.04 \pm 0.03 & 1.45 \pm 0.03\\
0.9 & 0.27 &  1.27 & 1.02 \pm 0.02 & 1.41 \pm 0.03\\
1 &   0.17    &  1.17 & 1.00             & 1.37 \pm 0.03\\ \hline\hline
\end{array}
$$
\renewcommand{\arraystretch}{1}
\addtolength{\arraycolsep}{-3pt}
\caption[]{The lowest-order coefficients $d_1^{V,A}$ and the resummed series
$M_\infty^{V,A}$ for $\eta_{V,A}$ in the $\overline{\rm MS}$ scheme as
functions of the ratio of pole masses
$m_c/m_b$. Note that the expressions diverge for $m_c\to 0$. For
$m_c=m_b$, $\eta_V\equiv 1$ due to charge conservation. Input parameter:
$-\beta_0\alpha_s(m_b) = 0.14$.}\label{tab:exkl}
\end{table}

To make an explicit comparison to inclusive decays possible, we have
presented the results in the form of an expansion
 in $\alpha_s(m_b)$ rather than in $\alpha_s(\sqrt{m_b m_c})$
which is more natural in exclusive decays. Because of this, our coefficients
are related to the ones given in  Ref.~\cite{NEU94}  by $d_1^{V,A}(\mu =
\sqrt{m_b m_c}) = d_1^{V,A}(\mu = m_b) + \ln m_c/m_b$.
Since the product of $\alpha_s$ and $M_\infty$ is scale-independent,
when a one-loop running coupling is used, we have
\begin{equation}
M^{V,A}_\infty[z,-\beta_0\alpha_s(\sqrt{m_bm_c})]\,=\,
\frac{\alpha_s(m_b)}{\alpha_s(\sqrt{m_b m_c})}\,
M^{V,A}_\infty[z,-\beta_0\alpha_s(m_b)]\,.
\end{equation}
Thus, for example, with $\alpha_s(m_b)/\alpha_s(\sqrt{m_bm_c}) = 0.82$ and
$M^{A}_\infty[0.3,-\beta_0\alpha_s(m_b)]=1.96$ from Table~\ref{tab:exkl},
we get $M^{A}_\infty[0.3,-\beta_0\alpha_s(\sqrt{m_bm_c})]= 1.59$.
Note that this number is rather large
and indicates that the higher-order corrections
 are important in the axial channel.

Where a comparison is possible, our results agree with the values obtained in
\cite{Nnew}. We obtain
\begin{equation}\label{PT-exclusive}
\eta_A = 0.943\pm 0.005 \pm 0.010 \pm  0.001,
\end{equation}
where the first error gives the estimated renormalon uncertainty\footnote{
We note that this uncertainty, although of order $\Lambda^2_{\rm QCD}/m^2$, is
numerically smaller than the estimate of explicit $1/m^2$ corrections,
see Sec.~4.2, which justifies the addition of these explicit corrections,
disregarding their potential ambiguity connected with renormalons.},
the second
one the uncertainty coming from $\alpha_s(m_Z)$, and the third one the
uncertainty in the input quark masses.

In the limit $m_c\to m_b$ the inclusive
decay rate thus equals (in perturbation theory)
\begin{equation}
\Gamma(B\to X_c e \bar \nu) = \frac{G_F^2 |V_{cb}|^2 (m_b-m_c)^5}{15\pi^3}\,
\left\{1-C_F \,\frac{\alpha_s(m_b)}{\pi}\,\frac{3}{4}\,M_\infty^A(a=1)
\right\},
\end{equation}
which implies the relation
\begin{equation}
M_\infty^{b\to c}(a=1) \equiv M_\infty^A(a=1)
\end{equation}
and provides a non-trivial check of our calculation. Comparing the
corresponding entries in
Table~\ref{tab:massive} and Table~\ref{tab:exkl} we indeed find agreement.

\section{Determination of $|V_{cb}|$}
\setcounter{equation}{0}
\subsection{Theoretical input parameters}
The main parameters we need in order to determine $|V_{cb}|$ are the b and the
c quark masses.
At present there seems to be no
general consensus on their values, and the existing estimates are often
controversial.  The masses extracted from the spectroscopy
of $\bar b b$ and $\bar c c$ mesons are usually given with very small
errors, see e.g.\ Ref.~\cite{VOL95}. However, the actual uncertainty is in this
case hidden in their  relation  to the running masses
at a certain hard scale, which we need in this paper. Thus, we prefer
to rely on a less accurate (as far as numbers are concerned) direct
determination of the $\overline{\rm MS}$ b quark mass from QCD sum rules
for lowest moments of $e^+e^-$ annihilation to heavy quarks
\cite{NOV78,SVZ}. Due to the smaller scales involved, such
estimates are less reliable for the c quark mass, so that we prefer to fix
$m_c$ by a different method, see below. In this paper we use the
following value for the $\overline{\rm MS}$ b quark mass
\cite{RRY81,NAR87}\footnote{
The number given in \cite{RRY81,NAR87} literally corresponds to the
Euclidian mass
$m_b^2(p^2=-m_b^2)$, but to the one-loop accuracy used in
Refs.~\cite{NOV78,SVZ} the difference between the Euclidian and the
$\overline{\rm MS}$ mass is negligible.}:
\begin{eqnarray}
  \overline{m}_b(\overline{m}_b) &=& (4.23 \pm 0.05)\,\mbox{\rm GeV.}
\label{eq:bquarkmass}
\end{eqnarray}
We point out that $\overline{m}_b$ was chosen as renormalization point
mainly in order to conform to the standard
choice in the literature,
in the same way as $\alpha_s$ is usually normalized by its value
at the Z boson mass. Actually in our approach the question of the ``natural''
scale does not appear, at least to the extent that the approximation of
summing higher order corrections due to vacuum polarization is good: all
results are explicitly scale-invariant for a one-loop running coupling.
In finite order perturbative calculations it may be more
appropriate to choose a lower renormalization scale in order to minimize the
size of uncalculated higher order terms.
{}From (\ref{eq:bquarkmass}) we get the pole mass\footnote{
Recall that any numerical value of the pole quark mass implies its
proper {\em definition}. Our
central value corresponds to the principal value prescription to sum the
perturbative series that relates the pole to the short-distance $\overline{\rm
MS}$ mass, the error comes from the $50\,$MeV uncertainty in
$\overline{m}_b(\overline{m}_b)$. The freedom in choosing the summation
prescription results in an
additional uncertainty of $m_b$ of order $100\,$MeV, which is exactly
cancelled by a corresponding uncertainty in the decay rate, see the
discussion in Sec.~3.3.}
\begin{eqnarray}\label{eq:bpolemass}
  m_b &=& (5.05 \pm 0.06 )\,\mbox{\rm GeV,}
\end{eqnarray}
where all radiative corrections of type
$\beta_0^n\alpha_s^{n+1}$ are resummed.
Very similar values for the b quark pole mass were proposed in
Refs.~\cite{NAR94,CHER95} and are also indicated by lattice calculations
\cite{mblatt}.

In order to fix the c quark mass, we make use of the fact that the
{\em difference\/} between the pole masses of two heavy quarks is free
from many ambiguities intrinsic in the mass parameters themselves and
can be determined to a good accuracy from the expansion
\begin{equation}\label{relate-bc}
 m_b-m_c =m_B -m_D +\frac{1}{2}\left(\frac{1}{m_b}-\frac{1}{m_c}\right)
\Big[\lambda_1+3\lambda_2\Big]+O(\alpha_s/m,1/m^2),
\end{equation}
 where $m_B$ and $m_D$ are the B and D meson masses, respectively; $\lambda_2$
is given by
\begin{equation}
 \lambda_2 \simeq \frac{1}{4}(m^2_{B^*}-m^2_B) \simeq 0.12\,
\mbox{\rm GeV}^2,
\end{equation}
and $-\lambda_1/(2m_b)$ is the kinetic energy of a heavy quark inside a B
meson. For $\lambda_1$ an estimate is available from QCD sum rules
\cite{BABR94},
$\lambda_1=-(0.6\pm 0.1)\,$GeV$^2$, which is however
strongly correlated with the  value of $\bar\Lambda =m_B-m_b$
and thus with the value of the b quark pole mass.
The estimate quoted in \cite{BABR94} is  the average of
$\lambda_1=-0.5\,$GeV$^2$ and $-0.7\,$GeV$^2$ obtained for
$\bar\Lambda = 400\,$MeV and $500\,$MeV, respectively. The resummation of
$\beta_0^n\alpha_s^{n+1}$ radiative corrections
 leads to a significant increase of the value of the
b quark pole mass, and thus to
a much lower value of $\bar\Lambda$ of order (200--300)$\,$MeV. Thus in
principle the value for $\lambda_1$ to be used in our analysis should be
obtained from a QCD sum rule using a small value of $\bar\Lambda$ and
including the resummation of running coupling effects, which is not
available. We have
calculated the BLM-type $\alpha_s^2$ correction to the simpler
sum rule for the leptonic decay constant $f_B$ in the
static limit \cite{BBBD}, which also enters the sum rule for $\lambda_1$
as normalization factor, and found that this correction is very small
(in other words: the BLM scale coincides with the ``naive''
hard scale). This may be accidental, however, and rather indicate that
other corrections are important. We have also checked that the
sum rule for $\lambda_1$ derived in \cite{BABR94}
becomes much less sensitive to the value of $\bar\Lambda$ if it
is small, and even with arbitrary $\bar\Lambda$ it is not possible to push
$-\lambda_1$ below (0.25--0.30)$\,$GeV$^2$. Lacking a BLM-improved sum
rule for $\lambda_1$, we consider $-(0.5\pm 0.2)\,$GeV$^2$ as a fair
estimate. With this value,
 one obtains\footnote{
Using our technique it is possible to estimate the uncertainty of
the relation (\ref{difference}) that is due to infra-red contributions
suppressed by three powers of the quark mass (as mentioned above,
the $1/m^2$ renormalon uncertainty is absent, at least to our
approximation). A simple calculation yields $\delta(m_b-m_c)
\simeq m_c/(4\pi|\beta_0|)\exp[5/2-3/(2|\beta_0|\alpha_s(m_c))]$,
which is of order (5--8)$\,$MeV.} from (\ref{relate-bc})
\begin{equation}\label{difference}
  m_b-m_c = (3.43 \pm 0.04)\,\mbox{\rm GeV}\,,
\end{equation}
assuming that $\alpha_s/m$ and $1/m^2$ corrections in (\ref{relate-bc})
are negligible.
Combining this result with the b quark pole mass in (\ref{eq:bpolemass})
we get the c quark pole mass
\begin{equation}\label{eq:cpolemass}
 m_c = (1.62\pm 0.07)\,\mbox{\rm GeV}
\end{equation}
and the running mass
\begin{equation}\label{eq:cquarkmass}
\overline{m}_c(\overline{m}_c) = (1.29\pm 0.06)\,\mbox{\rm GeV}\,.
\end{equation}
This value is consistent with determinations from
 QCD sum rules \cite{NOV78,SVZ}.

For completeness, we also give the corresponding values of
 the ``one-loop pole masses'' defined as
\begin{eqnarray}\label{pole1}
 m^{(1)}_{b,c} & = & \overline{m}_{b,c}(\overline{m}_{b,c})
   \Bigg\{1+C_F\frac{\alpha_s(\overline{m}_{b,c})}{\pi}\Bigg\}:\\
  m^{(1)}_{c} &=& (1.50\pm 0.06)\, \mbox{\rm GeV}\,,\nonumber\\
  m^{(1)}_{b} &=& (4.63\pm 0.05)\, \mbox{\rm GeV}\,.\label{set1-1}
\end{eqnarray}
Our values are in agreement with those quoted in Ref.~\cite{NAR94}.

\subsection{Extraction of $|V_{cb}|$}
Our discussion was so far restricted to the perturbative corrections
 to the b quark decay rate. In order to extract
$|V_{cb}|$ from the experimental data, we now have to specify the
non-perturbative corrections, which are suppressed by two powers of
the heavy quark masses. Summarizing
the results of Refs.~\cite{OPE,BABR94,BafflingBigi} we write
\begin{equation}
\Gamma(B\to X_c e \bar\nu) = \Gamma(B\to X_c e \bar\nu)_{pert}
\left(1+\frac{\delta^{NP}}{m_b^2}\right)\label{eq:nonpert}
\end{equation}
with $\delta^{NP} = -(1.05\pm 0.10)\,{\rm GeV}^2$, where the error comes from
the uncertainty in $\lambda_1$, cf.\ Sec.~4.1.

As for the exclusive decays, the experimentally interesting quantity is the
differential decay rate at zero recoil of the final state meson, which
depends on the form factor
\begin{equation}
{\cal F}(1) = \eta_A(1+\delta_{1/m^2}).
\end{equation}
The short-distance correction $\eta_A$ was already discussed in Sec.~3.5;
numerical values are given there and in Table~\ref{tab:inputOS} below. The
non-perturbative correction $\delta_{1/m^2}$ was estimated in
Refs.~\cite{NThUp}
as $-(5.5\pm 2.5)\%$ using $\lambda_1=-0.4$ GeV$^2$.\footnote{
An earlier estimate \cite{voodoo} was $-(8.5\pm 3)\%$ for
$\lambda_1=-0.54$ GeV$^2$.}

As experimental input we use the world average $B^0$ lifetime $\tau_{B^0}=
1.5\,$ps \cite{PData}, the most recent measurement of the $B^0$ semileptonic
branching ratio $B_{SL} = (10.9\pm 1.3)\%$ \cite{CLEObr}, and
$|V_{cb}{\cal F}(1)| = 0.0354 \pm 0.0027$ \cite{CLEOex}, where we have
rescaled the latter value to be compatible with $\tau_{B^0}=1.5\,$ps.
\begin{figure}[t]
\epsfxsize=0.5\textwidth
\centerline{\epsffile{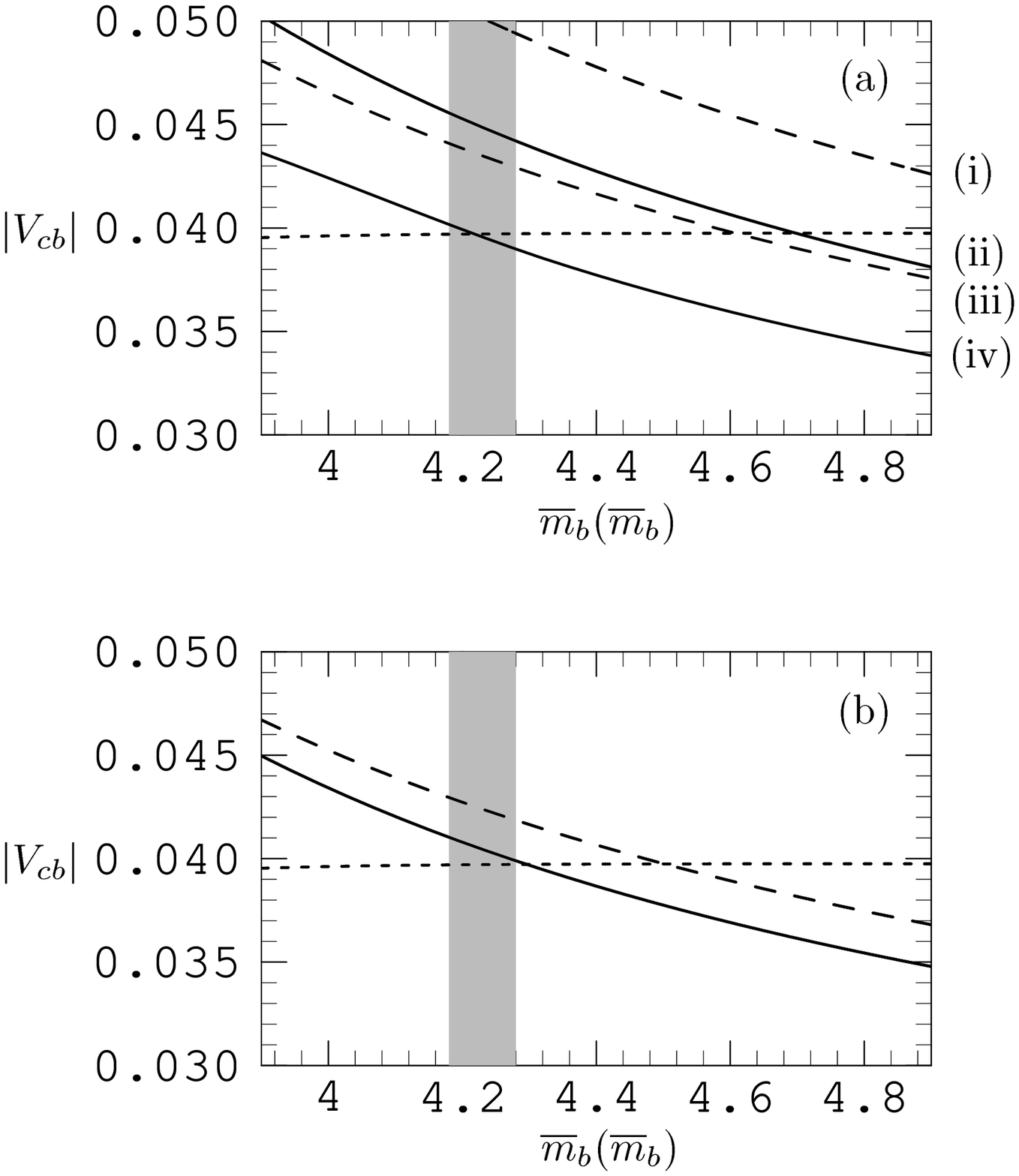}}
\caption[]{
The value of $|V_{cb}|$ extracted from the inclusive B meson
semileptonic decay rate to one-loop accuracy (a) [the curves (i) to (iv) are
explained in the text] and after resumming
$\beta_0^n\alpha_s^{n+1}$ radiative corrections (b) as a function of the
$\overline{\rm MS}$ b quark mass for fixed $\lambda_1 = -0.5\,$GeV$^2$.
The solid and long-dashed curves show the
predictions obtained by using the $\overline{\rm MS}$ and OS scheme,
respectively. The central value coming from exclusive decays is shown by
short dashes and the shaded area gives the interval of b quark mass values
suggested by QCD sum rules, Eq.\ (\ref{eq:bquarkmass}). Experimental input:
$\tau_{B^0} = 1.5\,$ps, $B_{SL}
=10.9\%$, $\alpha_s(m_Z)=0.117$.
}\label{fig:5}
\end{figure}
\begin{table}
\renewcommand{\arraystretch}{1.3}
\addtolength{\arraycolsep}{3pt}
$$
\begin{array}{cl|lllll}
& \overline{m}_b(\overline{m}_b) & 4.0 & 4.1 & 4.2 & 4.3 & 4.4\\
& \multicolumn{1}{r|}{m_b}        & 4.80^{+(9)}_{-(9)} & 4.91^{+(9)}_{-(9)} &
5.02^{+(9)}_{-(9)} & 5.13^{+(9)}_{-(9)} & 5.24^{+(9)}_{-(9)}\\
\overline{m}_c(\overline{m}_c) & m_c \\\hline\hline
& \\[-14pt]
1.18 & 1.47^{-(6)}_{+(1)} & 0.045^{-(3)}_{+(3)} & 0.042^{-(3)}_{+(2)} &
0.039^{-(2)}_{+(2)} & 0.037^{-(2)}_{+(2)} & 0.034^{-(2)}_{+(2)}\\[4pt]
1.22 & 1.52^{-(5)}_{+(0)} & 0.046^{-(3)}_{+(3)} & 0.043^{-(3)}_{+(2)} &
0.040^{-(2)}_{+(2)} & 0.037^{-(2)}_{+(2)} & 0.035^{-(2)}_{+(2)}\\[4pt]
%1.24 & 1.55^{-(5)}_{+(0)} & 0.047^{-(3)}_{+(3)} & 0.043^{-(3)}_{+(2)} &
%0.040^{-(2)}_{+(2)} & 0.038^{-(2)}_{+(2)} & 0.035^{-(2)}_{+(2)}\\[4pt]
1.26 & 1.58^{-(4)}_{+(0)} & 0.047^{-(3)}_{+(3)} & 0.044^{-(3)}_{+(2)} &
0.041^{-(2)}_{+(2)} & 0.038^{-(2)}_{+(2)} & 0.035^{-(2)}_{+(2)}\\[4pt]
%1.28 & 1.60^{-(4)}_{+(0)} & 0.048^{-(3)}_{+(3)} & 0.044^{-(3)}_{+(2)} &
%0.041^{-(3)}_{+(2)} & 0.038^{-(2)}_{+(2)} & 0.036^{-(2)}_{+(2)}\\[4pt]
1.30 & 1.63^{-(4)}_{+(0)} & 0.048^{-(3)}_{+(3)} & 0.045^{-(3)}_{+(2)} &
0.041^{-(3)}_{+(2)} & 0.039^{-(2)}_{+(2)} & 0.036^{-(2)}_{+(2)}\\[4pt]
1.34 & 1.68^{-(3)}_{-(1)} & 0.049^{-(3)}_{+(3)} & 0.046 ^{-(3)}_{+(3)} &
0.042^{-(3)}_{+(2)} & 0.039^{-(2)}_{+(2)} & 0.037^{-(2)}_{+(2)}\\[4pt]
\hline\hline
\end{array}
$$
\caption[]{$|V_{cb}|$ calculated in the $\overline{\rm MS}$ scheme for
a wide range of
quark masses. The central value is obtained for $\alpha_s(m_Z)=0.117$, the
upper (lower) number in brackets gives the shift of the last digit if
$\alpha_s$ is put to $0.123$ ($0.111$). The masses are given in GeV.
Experimental
input: $B_{SL} = 10.9$\%, $\tau_{B^0} = 1.5\,$ps.}\label{tab:Vcb1}
$$
\begin{array}{cc|lllll}
& \multicolumn{1}{r|}{\overline{m}_b(\overline{m}_b)} & 4.0 & 4.1 & 4.2 & 4
.3 & 4.4\\
& \multicolumn{1}{r|}{m_b} & 4.80^{+(9)}_{-(9)} & 4.91^{+(9)}_{-(9)} &
5.02^{+(9)}_{-(9)} & 5.13^{+(9)}_{-(9)} & 5.24^{+(9)}_{-(9)}\\
\lambda_1\,[{\rm GeV}^2] & \multicolumn{1}{l|}{m_b-m_c} \\\hline\hline
& \\[-14pt]
-0.3 & 3.39 & 0.0441^{-(7)}_{+(7)} & 0.0426^{-(7)}_{+(7)} &
0.0414^{-(7)}_{+(7)} & 0.0402^{-(7)}_{+(7)} & 0.0392^{-(6)}_{+(6)}\\[4pt]
-0.4 & 3.41 & 0.0438^{-(7)}_{+(7)} & 0.0423^{-(7)}_{+(7)} &
0.0411^{-(6)}_{+(7)} & 0.0399^{-(6)}_{+(6)} & 0.0389^{-(6)}_{+(6)}\\[4pt]
-0.5 & 3.43 & 0.0434^{-(6)}_{+(7)} & 0.0420^{-(6)}_{+(6)} &
0.0408^{-(6)}_{+(6)} & 0.0397^{-(6)}_{+(6)} & 0.0387^{-(6)}_{+(6)}\\[4pt]
-0.6 & 3.46 & 0.0431^{-(6)}_{+(6)} & 0.0417^{-(6)}_{+(6)} &
0.0405^{-(6)}_{+(6)} & 0.0394^{-(6)}_{+(6)} & 0.0384^{-(6)}_{+(6)}\\[4pt]
-0.7 & 3.48 & 0.0427^{-(6)}_{+(6)} & 0.0414^{-(6)}_{+(6)} &
0.0402^{-(6)}_{+(6)} & 0.0391^{-(6)}_{+(6)} & 0.0382^{-(5)}_{+(5)}\\[4pt]
\hline\hline
\end{array}
$$
\renewcommand{\arraystretch}{1}
\addtolength{\arraycolsep}{-3pt}
\caption[]{$|V_{cb}|$ calculated in the $\overline{\rm MS}$ scheme
in dependence on
the b quark mass and $\lambda_1$. Notations and experimental
input as in the previous table.}\label{tab:Vcb2}
\end{table}

Our results are summarized in Fig.~\ref{fig:5} and in Tables~\ref{tab:Vcb1}
and \ref{tab:Vcb2}, where
we give $|V_{cb}|$ as function of the running b quark mass. The determination
with one-loop radiative corrections is plotted
in Fig.~\ref{fig:5}(a) and using resummation in Fig.~\ref{fig:5}(b).
The solid line represents the result from the $\overline{\rm MS}$,
the long dashes from the OS calculation. The short dashes give $|V_{cb}|$ from
exclusive decays. The shaded areas illustrate the range of b quark mass
values from Eq.\ (\ref{eq:bquarkmass}). The curves
in Fig.~\ref{fig:5}(b)
are obtained for
$\lambda_1 = -0.5\,$GeV$^2$, which corresponds roughly to a fixed difference
of the pole masses of $m_b-m_c = 3.43\,$GeV.

The two sets of curves in Fig.~\ref{fig:5}(a) are obtained with two
different ways to specify the c quark mass.
The first way is to use exactly the same short-distance
c quark mass as in Fig.~\ref{fig:5}(b): the curves
labelled (i) and (ii) are obtained using the constraint (\ref{relate-bc}) with
 {\em resummed} pole masses,
i.e.\ fixing $m_b-m_c=3.43\,$GeV, and calculating
$m_c^{(1)}$ and $\overline{m}_c
(\overline{m}_c)$, respectively, for a given value of $\overline{m}_b
(\overline{m}_b)$.
Thus, since all short-distance input parameters are chosen in precisely the
same way as in Fig.~\ref{fig:5}(b), the difference between the predictions
for $|V_{cb}|$ is entirely the effect of resummation.

Although this choice is presumably the most clear way to show the effect
of resummation of running coupling effects, it also may  be slightly
misleading. Indeed, the large difference in values of $|V_{cb}|$ between the
one-loop and resummed formulas is mainly due to the fact that the one-loop
pole masses defined in this way do not satisfy Eq.~(\ref{relate-bc}):
 $m_b^{(1)}-m_c^{(1)}=3.13\,$GeV.
Thus we also try another choice, calculating the c quark mass
by enforcing the constraint (\ref{relate-bc}) expressed in terms of
{\em one-loop} pole masses, so
that $m_b^{(1)}-m_c^{(1)}=3.43\,$GeV. The results are shown in the form
of curves (iii) and (iv). This choice is less instructive
as far as the comparison between one-loop and resummed results is
concerned, but is probably more attractive phenomenologically.
On the other hand, we note that with this choice
the value of the short-distance c quark mass becomes very low,
$\overline{m}_c(\overline{m}_c)=0.98\,$GeV [$m_c^{(1)} =
1.18\,$GeV] for
$\overline{m}_b(\overline{m}_b)=4.23\,$GeV, which is hardly
 consistent with the QCD sum rules for the charmonium system.
The large difference in the resulting
values for $|V_{cb}|$  shows the dilemma all strict one-loop
calculations are inevitably confronted with: it is impossible to relate the
three independently determined input parameters
$\overline{m}_b(\overline{m}_b)$, $\overline{m}_c(\overline{m}_c)$ and
$\lambda_1$ to each other within the errors by one-loop equations, although
all of them provide valid phenomenological input. It is only after inclusion
of higher order perturbative corrections in Eq.~(\ref{relate-bc}) in form
of the resummed pole masses that the
three values appear to be consistent with each other, and it
may be considered as a serious argument in favour
of BLM-improved perturbation theory in semileptonic inclusive decays
 that the central value
(but not the error bars, see below) of $|V_{cb}|$ is independent of the
choice of a particular subset.

The effect of resummation is clearly visible in Fig.~\ref{fig:5}(b):
first, we find a considerably reduced scheme dependence,
i.e.\ the difference between the solid and the long-dashed curves is much
smaller in  Fig.~\ref{fig:5}(b) than in Fig.~\ref{fig:5}(a).
Secondly, we observe good agreement between $|V_{cb}|$ from exclusive and
inclusive decays obtained with resummation, which otherwise is only
achieved for
either an unreasonably large b quark mass or a very small c quark mass.

It has been proposed in Ref.~\cite{voodoo} that in inclusive decays the
dependence on the quark masses is significantly reduced, if the charm and
bottom masses are not varied independently, but related to each other
by Eq.~(\ref{relate-bc}).
To study this question, we give tables of numerical values for
$|V_{cb}|$, choosing as independent input parameters either
 the running b and c quark masses
(Table~\ref{tab:Vcb1}), or  the running b quark mass and $\lambda_1$,
which specifies the difference between the pole masses
(Table~\ref{tab:Vcb2}).
It is clearly seen that, although the central values for $|V_{cb}|$ are
nearly the same, the latter choice is preferable, since the inclusive decay
rate is very sensitive to the mass difference $m_b-m_c$, and already a
very modest accuracy in $\lambda_1$ in fact constrains
$m_b-m_c$ more precisely\footnote{Even if we
abandoned the determination of $\lambda_1$ in Ref.~\cite{BABR94} completely
and only put the constraint $-0.7\,{\rm GeV}^2 \leq \lambda_1 \leq 0\,
{\rm GeV}^2$, which corresponds to $3.34\,{\rm GeV}\leq m_b-m_c\leq 3.48\,
{\rm GeV}$, $|V_{cb}|$ would change by at most $0.0014$.}
than any direct determination. We conclude that
$\lambda_1$ (or equivalently $m_b-m_c$) is a better theoretical input
 parameter than the c quark mass itself, in agreement with
the discussion in Ref.~\cite{voodoo}. In addition we find that
 the dependence on $\alpha_s(m_Z)$ is strongly reduced, too. We emphasize,
however, that our choice of input parameters only serves to reduce the
sensitivity of $|V_{cb}|$ on uncertainties in the input parameters and
that the central value is independent of that choice.

It might also be useful to compare the results obtained
from the resummed formulas with those obtained in the usual BLM approximation,
where of the whole series of corrections generated by the running of the
coupling only the $\alpha_s^2\beta_0$ term is taken into account.
Again, the major subtlety comes from the necessity to specify the c quark
mass. It turns out that the perturbative series  that relates the
c quark pole mass to the $\overline{\rm MS}$ mass starts to diverge already
in second order \cite{BB94b}. Because of the divergence, it is {\em not}
justified to cut the series at that order, although in this case the BLM
correction gives an excellent approximation to the exact two-loop result.
For example, starting from
$\overline{m}_c(\overline{m}_c)=1.26\,$GeV one gets for the BLM pole mass
$m_c^{BLM}= 1.75$ GeV, which is significantly
{\em larger} than the resummed result $m_c=1.58$ GeV. In other words,
the BLM approximation underestimates the scale of the
coupling and thus overestimates the radiative correction.
A comparison between the BLM and the resummed results is however possible if
one starts from the $\overline{\rm MS}$ b quark
mass, calculates the b quark pole mass in the BLM approximation, and then
fixes the c quark pole mass from the heavy-quark expansion (\ref{relate-bc}).
Taking for definiteness $\overline{m}_b(\overline{m}_b)=4.23\,$GeV and
$\lambda_1 = -0.5\,$GeV$^2$, we find in this way in the BLM approximation
$|V_{cb}|=0.0381$ ($|V_{cb}|=0.0424$) in the $\overline{\rm MS}$ (OS) scheme,
respectively, compared with $|V_{cb}|=0.0404$ ($|V_{cb}|=0.0424$) with resummed
formulas. The perfect agreement between the BLM and resummed formulas in
the OS scheme is probably accidental. In the $\overline{\rm MS}$ (OS) scheme
the BLM result is very close to the one-loop calculation, see Fig.~\ref{fig:5}.
In both cases, the numerical effect of higher order corrections is
small and of the order of 5\%.
We wish to emphasize once more, however, that this comparison is only
possible if the c quark mass is obtained in a very particular way
and it is only after resummation that we are able to get a selfconsistent
description in terms of short-distance parameters.

{}From the combined evidence of Fig.~\ref{fig:5} and Tables~\ref{tab:Vcb1}
and \ref{tab:Vcb2} we extract the following results:
\begin{eqnarray}
|V_{cb}|^{\overline{\rm\scriptsize MS}} & = & (0.0404\pm 0.0006 \pm 0.0006
\pm 0.0006)\,
\left(\frac{B_{SL}}{10.9\%}\right)^{1/2}
\left(\frac{1.5\,{\rm ps}}{\tau_{B^0}}\right)^{1/2},
\label{eq:resMS}\\
|V_{cb}|^{{\rm\scriptsize OS}} & = & (0.0424\pm 0.0004 \pm 0.0007
\pm 0.0005)\,
\left(\frac{B_{SL}}{10.9\%}\right)^{1/2}
\left(\frac{1.5\,{\rm ps}}{\tau_{B^0}}\right)^{1/2},
\label{eq:resOS}
\end{eqnarray}
where the first error comes from the uncertainty in $\alpha_s(m_Z)=0.117\pm
0.006$, the second one from the uncertainty of the b quark mass, Eq.\
(\ref{eq:bquarkmass}), and the third one from the uncertainty in $\lambda_1=
-(0.5\pm 0.2)\,$GeV$^2$.
      For the reader's convenience, we collect all other parameters
      in Table~\ref{tab:inputOS} and  Table~\ref{tab:inputMS}.
\begin{table}
\addtolength{\arraycolsep}{2pt}
\renewcommand{\arraystretch}{1.2}
$$
\begin{array}{cc|cccccc}
\alpha_s(m_Z) & \Lambda_{\overline{\rm\scriptsize MS}}^{(4)}\,[{\rm MeV}]
& m_b\,[{\rm GeV}] & m_c\,[{\rm GeV}] & \alpha_s(m_b) & \alpha_s(m_c) &
M_\infty^{b\to c}[a,-\beta_0^{(4)}\alpha_s(m_b)] & \eta_A\\ \hline\hline
0.111 & 223 &4.96 &1.52 &0.190 &0.291 &2.07 & 0.954\\
0.117 & 308 &5.05 &1.62 &0.208 &0.333 &2.07 & 0.943\\
0.123 & 411 &5.15 &1.72 &0.228 &0.384 &2.02 & 0.938\\ \hline\hline
\end{array}
$$
\caption[]{
Input parameters in the calculation of the
 resummed B decay rate in the OS scheme and values of $\eta_A$ entering the
exclusive decay rate.
}\label{tab:inputOS}
$$
\begin{array}{c|ccccccc}
\alpha_s(m_Z) & \sqrt{\bar a} & \alpha_s(\overline{m}_b) &
\alpha_s(\overline{m}_c) & \overline{M}_\infty^b(\overline{m}_b) &
\overline{M}_\infty^c(\overline{m}_c) &
M_\infty^{b\to c}[\bar a,-\beta_0^{(4)}\alpha_s(\overline{m}_b)] &
\overline{M}_\infty^{b\to c}\\ \hline\hline
0.111 & 0.287 &0.199 &0.326 &2.04 &1.82 &2.06 & 2.18\\
0.117 & 0.306 &0.220 &0.383 &2.08 &1.55 &1.97 & 2.40\\
0.123 & 0.329 &0.244 &0.451 &2.09 &1.22 &1.82 & 2.68
\\ \hline\hline
\end{array}
$$
\caption[]{
Input parameters in the calculation of the
 resummed B decay rate in the $\overline{\rm MS}$ scheme.
}\label{tab:inputMS}
\addtolength{\arraycolsep}{-2pt}
\renewcommand{\arraystretch}{1}
\end{table}

Comparing  Eqs.\ (\ref{eq:resMS}) and (\ref{eq:resOS}), it is clear that the
uncertainty of our calculation is dominated by scheme dependence, which is
simply the effect of higher-order radiative corrections not related to
the running of the QCD coupling and which we miss in our approximation.
A different way to estimate
unknown higher-order corrections is to consider the scale dependence of
the results in the $\overline{\rm MS}$ scheme. This
scale dependence is due to two-loop running effects in $\alpha_s$ and
terms of order $(\alpha_sC_F)^2$, which are beyond the accuracy
of our approach.
We have checked that after renormalization group improvement of
$\ln(\mu^2/m_b^2)$ corrections (i.e.\ using the exponentiated form of
the b quark mass scale dependence as in Eq.~(\ref{eq:RGmass}))
the variation of $|V_{cb}|$ with the
renormalization scale $\mu$ within $m_b/2 \leq \mu \leq 2 m_b$ is of order
$\pm 0.001$, i.e.\ of the same order as the difference between the OS and
$\overline{\rm MS}$ calculations.
This may be taken as an estimate of the accuracy of the resummation,
beyond which an explicit calculation of higher-order corrections is
necessary.
Combining the errors, we get
\begin{equation}
\left(\tau_{B^0}/{1.5\,{\rm ps}}\right)^{1/2}
|V_{cb}|_{incl} = 0.041\pm 0.002 \pm 0.002,
\end{equation}
which is our final result.
The first error shows the theoretical uncertainty
and the second error comes from the experimental semileptonic
branching ratio.
The theoretical error is dominated by the
uncalculated exact $\alpha_s^2$ correction to the decay rate.
All large corrections coming from the running of the strong coupling
either cancel after proper treatment of the infra-red regions or are
cast into the redefinition of the scale in the coupling or, equivalently,
into the  ``M-factors''. The numerical significance of these M-factors depends
on the way how the input parameters are chosen, and can be minimized
by using the constraint (\ref{relate-bc}) following from the
heavy quark expansion.

In turn, we get from the exclusive decays
\begin{equation}\label{exx}
\left(\tau_{B^0}/{1.5\,{\rm ps}}\right)^{1/2}
|V_{cb}|_{excl} = 0.040\pm 0.001 \pm 0.003,
\end{equation}
where the first error is the theoretical uncertainty\footnote{
The theoretical error indicated in (\ref{exx}) does not include
neither possible perturbative corrections, not related to
running of the coupling, nor the uncertainty in
 $\lambda_1$ which contributes to the overall $1/m^2$ correction.}
 and the
second one the experimental error of the decay rate.

 At present the experimental errors are roughly the same for both
the exclusive and the inclusive determinations.
Actually the two approaches are complementary to each other:
the inclusive decays
offer a better opportunity to reduce the experimental errors,
whereas the exclusive decay rates
are inevitably plagued with large statistical errors
from a measurement near the edge of phase space. On the other hand,
in inclusive decays, the
theoretical predictions are
more sensitive to  errors in the quark masses.
It is encouraging that even now, with moderate experimental accuracy
and an improvable accuracy of the theoretical input, both methods lead to
very similar results.

\begin{table}
\renewcommand{\arraystretch}{1.2}
\begin{tabular}{l|llllll}
& \multicolumn{2}{c}{This paper$^{(a)}$} & Ref.~\cite{LS}$^{(b)}$ &
Ref.~\cite{BU}$^{(b)}$ & Ref.~\cite{BN94}$^{(b)}$ &
Ref.~\cite{voodoo}$^{(a)}$\\ \hline\hline
Scheme & $\overline{\rm MS}$ & OS & OS & OS & $\overline{\rm MS}$ & OS\\
$m_c\,$[GeV] & $1.26^{(c)}$ & $1.62^{(d)}$ & -- & $1.57^{(d)}$ & $1.35^{(c)}$
& $1.3^{(d)}$\\
$m_b\,$[GeV] & $4.23^{(c)}$ & $5.05^{(d)}$ & -- & $4.96^{(d)}$ & $4.6^{(c)}$
& $4.8^{(d)}$\\
$|V_{cb}|$ & $0.040(1)$ & $0.042(1)$ & $0.046(8)$ &
$\approx 0.042$ & $0.036(3)$ & $0.042$ \\ \hline\hline
\end{tabular}
\renewcommand{\arraystretch}{1}
\makebox[2cm]{ }\\[3pt]
{\footnotesize
\begin{tabular}{l@{\hspace{1cm}}l}
  ${}^{(a)}$ $m_b$ from QCD sum rules,
  $m_c$ from Eq.\ (\ref{relate-bc}).  & ${}^{(c)}$
  Running mass normalized at $\overline{m}$.\\ ${}^{(b)}$ $m_c$ from
  $B(D\to X e\bar\nu)$, $m_b$ from Eq.\ (\ref{relate-bc}).  & ${}^{(d)}$
  Pole mass.
\end{tabular}
}
\caption[]{$|V_{cb}|$ from inclusive decays obtained in previous
analyses. If necessary, the quoted numbers are rescaled to be compatible
with $\tau_{B^0} = 1.5\,$ps, $B_{SL} = 10.9$\%. The quark masses are the
central values used in the papers.}\label{tab:comparison}
\end{table}

\section{Conclusions}
We have carried out a detailed analysis of the radiative corrections
to inclusive semileptonic B decays that originate from the running
of the strong coupling. Independent of the actual accuracy of this
approximation, our results clearly indicate that the series of radiative
corrections in the OS scheme diverges, starting already in low orders.
Thus determinations of $|V_{cb}|$ from inclusive decays using the
pole masses of b and c quarks as input parameters are plagued with a
numerically large uncertainty. The accuracy cannot
be improved by choosing a different scheme or a lower scale in the
running coupling.

However, the problem is spurious and entirely due to the use
of a bad input parameter --- the pole mass --- which imports infra-red
contributions at the level of $O(1/m_b)$ corrections. It can be shown
\cite{BIG94,BBZ94} that these corrections are absent in the inclusive decay
rates, and therefore using pole masses in the tree-level decay rate
induces large radiative corrections of infra-red origin
simply in order to cancel infra-red effects hidden in the definition
of the mass parameter.

We demonstrate that the behaviour of the perturbative series is indeed
drastically improved by using the $\overline{\rm MS}$ mass instead.
The calculation
in the OS scheme can be saved, if the pole mass is defined by a certain
non-perturbative prescription in its relation to the short-distance mass
(or some physical quantity from which it is determined),
and if the same prescription is used to sum the series of radiative
corrections to the decay widths.
This essentially implies a rearrangement of radiative corrections in two
pieces, hiding part of them in the tree-level phase space, and in practice
involves considerable cancellations already in low orders.

We carry out a detailed analysis of inclusive decay rates
with resummation of corrections induced by one-loop running of
the coupling and find a good agreement
between values of $|V_{cb}|$ extracted from inclusive and exclusive
decays. The comparison of our results with earlier calculations
is presented in Table~\ref{tab:comparison}. In general, we find
agreement within the errors. It should be emphasized, however, that
predictions for
inclusive decays depend rather strongly on the quark masses, and
effects of the resummation can be masked by using different input
values. We believe that one advantage of our approach lies in using
well-defined mass parameters, which can (in principle) be extracted
from experimental data with high accuracy.

The accuracy of our predictions is limited by the
unknown accuracy of the resummation of BLM-type radiative corrections.
The incompleteness of this procedure is at least partially indicated
by the scheme dependence of the result for $|V_{cb}|$, which is
of order 5\%. To reduce this remaining error it will be necessary
to incorporate exact $\alpha_s^2$ corrections to
the decay rates. One could try to reduce radiative corrections
by expressing the decay widths in terms of masses renormalized
at smaller scales,
and adjust their values in order to reproduce the bulk
of available data on heavy hadrons in the framework of one-loop
calculations. This approach can be phenomenologically successful
inasmuch as the structure of higher-order radiative corrections
is similar in various processes, which is natural to assume, but
difficult to control theoretically.

On the other hand, with a restriction to the level of accuracy
of order 5\%  for  $|V_{cb}|$,
we find that the theoretical calculation can be justified, and
only a moderate precision is required for input parameters such as
quark masses. In particular, it is sufficient to know the running b quark
mass to an accuracy of order (50--100)$\,$MeV, and the difference in pole
masses of b and c quarks to $\sim 50\,$MeV, which corresponds to
the uncertainty of the kinetic energy of the b quark inside
the B meson of order $0.2\,$GeV$^2$. We think that it is possible to
achieve this kind of accuracy, and in fact even to improve it.

To summarize, we conclude that inclusive decays of B mesons
provide a valid source of information on $|V_{cb}|$ with a
present theoretical accuracy of order 5\%, which presumably can be
improved in the future.\\

\noindent {\bf Acknowledgements}:
M.$\,$B.\ and V.$\,$B.\ gratefully acknowledge the hospitality of the
CERN theory
group, where this work was completed. M.$\,$B.\ thanks I. Rothstein for
interesting discussions and the Alexander von Humboldt foundation
for financial support. V.$\,$B.\ is grateful to
V.$\,$ Chernyak and N.G.$\,$ Uraltsev for
critical remarks. After completion of this work, Ref.~\cite{Unew} appeared,
which partially overlaps with our discussion.

%\clearpage

%\section*{Appendices}

\appendix
\setcounter{equation}{0}
\renewcommand{\theequation}{\Alph{section}.\arabic{equation}}
\renewcommand{\thetable}{\Alph{table}}
\setcounter{table}{0}

\section{Radiative corrections to inclusive decays}

In this appendix we give an explicit expression for the one-loop radiative
correction
to the total inclusive decay width. We use the on-shell
renormalization scheme, where for finite
gluon mass the wave-function renormalization constant is given by (in
dimensional regularization in $D$ dimensions):
\begin{eqnarray}
Z_{2F} & = & 1 + C_F\,\frac{\alpha_s}{4\pi}\left\{ \frac{2}{D-4} + \gamma_E -
\ln 4 \pi + \ln\,\frac{m_b^2}{\mu^2} -4 - 3 y + \left( \frac{3}{2} \, y^2 -
2 \right) \ln y\right.\nonumber\\
& & \left.\phantom{1 + C_F\,\frac{g_s^2}{(4\pi)^2}\left\{ \right.}{} -
\frac{3(y^2-2y-4)}{2\sqrt{1-4/y}}\,\ln\,
\frac{1+\sqrt{1-4/y}}{1-\sqrt{1-4/y}}\, \right\}.
\end{eqnarray}
For a massive quark's self-energy $\Sigma$ we impose the
renormalization condition $\Sigma^R(m^2) \stackrel{!}{=} 0$.
Throughout the appendix, we use the notations $y = \lambda^2/m_b^2$ and
$ a = m_c^2/m_b^2$.

The function $d_0(a,\lambda^2)$ can be written as the sum
of contributions of real and virtual gluon emission, corresponding to
different imaginary parts of the diagrams in Fig.~\ref{fig:1}:
\begin{equation}
 f_1(a) g_0(a)d_0(a,\lambda^2) = -\left[D^{virt}(a,y) +
                      \theta(m_b-m_c-\lambda)D^{brems}(a,y)\right]
\end{equation}
(recall that $f_1$ is the tree-level phase space factor defined in
(\ref{def:f1}) and $g_0(a)$ the one-loop correction for zero gluon mass).
The subscripts accompanying the $D$'s specify
the diagrams in Fig.~\ref{fig:2}, from which they are obtained,
so that
\begin{equation}
      D= D_{I}+D_{II}+D_{III}+D_{III}^{\dagger}+D_{IV}.
\end{equation}

\subsection{The decay $b\to c e \bar\nu$}

For a massive quark in the final state, we did not succeed in obtaining
an analytic expression. Below we give the renormalized contributions
of the single diagrams, in Feynman gauge,
expressed in terms of at most two-dimensional integrals:
\begin{eqnarray}
D_{I} & = & f_1(a)\left\{-1 - \frac{3\,y}{4} +
\frac{3(4+2\,y-y^2)}{8\sqrt{1-4/y}}\, \ln\,
\frac{1+\sqrt{1-4/y}}{1-\sqrt{1-4/y}} +
\left(\frac{3}{8}\,y^2-\frac{1}{2}\right)\,\ln y\right\}\!,\hskip10pt\\
D_{II}^{virt} & = & \left. D_{II} + f_1(a)\, \frac{\ln a}{4}
\right|_{y\to y/a},\\
D_{II}^{brems} & = & \label{eq:derAnfang}
\hskip-15pt\int\limits_{(\sqrt{a}+\sqrt{y})^2}^1
\hskip-15ptdx\,
f_1(x)\, w(a,x,y)\, \frac{a^2+x^2-6 a x-y (a+x)}{4x^2(x-a)^2},\\
\lefteqn{\hskip-1.2cm D_{III}^{virt}+D_{III}^{\dagger\,virt} = \hskip-10pt\int
\limits_0^{(1-\sqrt{a})^2}\hskip-10ptdx\int\limits_0^1 dz\,w(1,a,x)\left[
-2xy+\frac{\ln a}{2a} \left(-4a+2y+2a^2+y^2+2a^3+2 a x\right.\right.}
\nonumber\\
& & \left. + 2 a y+2 x y-4a^2x-4a^2y+2ax^2-ay^2-3xy^2-4x^2y+2axy\right)
\nonumber\\
& & {}+ w(1,a,x)\left(3x + 2y\right) \ln\,
\frac{1+\sqrt{1-4a/(1+a-x)^2}}{1-\sqrt{1-4a/(1+a-x)^2}} + \frac{y}{2}\,\sqrt{
1-\frac{4}{y}}\nonumber\\
& & {}\times(-2-y+2a^2-4x^2+2ax+ay-3xy)\, \ln \,\frac{1+\sqrt{1-4/y}}{1-
\sqrt{1-4/y}} + \frac{y}{2a}\,\sqrt{1-\frac{4a}{y}}\nonumber\\
& & {}\times(2+2x+y-2a^2-4x^2-ay-3xy) \ln\,\frac{1+\sqrt{1-4a/y}}{1-
\sqrt{1-4a/y}} + \frac{y\ln y}{2a}\, (-2+2a\nonumber\\
& & {}-2x-y+2a^2-2a^3+4x^2-4ax+2ay-2a^2x-a^2y+4ax^2+3xy+3axy) \nonumber\\
& & {}-\frac{1}{A}\, (1-a-a^2+a^3-3x^2+2x^3+y+4ax-3ax^2-2ay+a^2y-2xy
\nonumber\\
& & \left. {}+2xy^2+4x^2y-2axy) \left(\ln A-\ln y +
\frac{1}{\sqrt{1-4A/y}}\, \ln \,\frac{1+\sqrt{1-4A/y}}{1-\sqrt{1-4A/y}}
\right)\right],\nonumber\\[-10pt]
& &\\
\lefteqn{\hskip-1.2cmD_{III}^{brems}+D_{III}^{\dagger\,brems} = \hskip-15pt\int
\limits_{(\sqrt{a}+\sqrt{y})^2}^1\hskip-15ptdx \hskip-10pt\int
\limits_0^{(1-\sqrt{x})^2}\hskip-10pt
dz\,\left[\frac{w(1,x,z)\,w(a,x,y)}{x(a-x)}\,\left( 2+a-x+y+2z+x^2-4z^2
\right.\right.}\nonumber\\
& & \left. {}-3ax+az-xy-xz-3yz \right) + \frac{2}{a-x}\,\left(
1-x+y+x^2-3z^2+2z^3-2ax\right.\nonumber\\
& & \left. {}-ay+4az-xy-2yz+a^2x-a^2z-az^2-2xz^2+4yz^2-2y^2z+axy\right.
\nonumber\\
& & \left. \left. {}+axz-2xyz\right)\,\ln {{ w(1,x,z)\,w(a,x,y) +
       \left( 1 - z + x \right) \,\left( a - x - y \right)  + 2\,x\,y}\over
     {-w(1,x,z)\,w(a,x,y) +
       \left( 1 - z + x \right) \,\left( a - x - y \right)  + 2\,x\,y}}\right],
\nonumber\\[-10pt]
& &\\
D_{IV}^{brems} & = & \hskip-12pt\int\limits_a^{(1-\sqrt{y})^2}\hskip-10pt dx\,
f_1(a/x)\, w(1,x,y)\, \frac{x^2\{1+x^2-6 x-y (1+x)\}}{4(1-x)^2},
\label{eq:vomEnde}
\end{eqnarray}
where we have used
\begin{equation}
w(x,y,z) = (x^2+y^2+z^2-2xy-2xz-2yz)^{1/2},\quad A = 1-z+az-x(1-z)z.
\end{equation}
\begin{table}
\addtolength{\arraycolsep}{3pt}
\renewcommand{\arraystretch}{1.3}
$$
\begin{array}{l|lllllllllll}
\sqrt{a} & 0 & 0.1 & 0.2 & 0.3 & 0.4 & 0.5 & 0.6 & 0.7 & 0.8 & 0.9 & 0.99
\\ \hline\hline
\kappa_2 & \frac{4321}{432}-\frac{5\pi^2}{9} & 4.53 & 4.57 & 4.62 & 4.71 &
4.89 & 5.25 & 5.89 & 6.96 & 8.64 & 10.83\\
\kappa_3 & \frac{667133}{14700} - \frac{8\pi^2}{3} & 19.1 & 19.2 & 19.4 &
19.6 & 19.9 & 20.6 & 22.1 & \multicolumn{1}{c}{-} & \multicolumn{1}{c}{-} &
\multicolumn{1}{c}{-}\\ \hline\hline
\end{array}
$$
\addtolength{\arraycolsep}{-3pt}
\renewcommand{\arraystretch}{1}
\caption[]{Coefficient-functions in Eq.\ (\ref{largeexpand}).}\label{tab:large}
\end{table}

The integrals have been further evaluated numerically. In doing so
it is convenient to use the following expansion for large $y$:
\begin{equation}\label{largeexpand}
D^{virt}\stackrel{y\to\infty}{=} \sum\limits_{n=1}^3 [\kappa_n(a)
+ \zeta_n(a) \ln y]\,\frac{f_1(a)}{y^n} + O(\ln y/y^4)
\end{equation}
with the coefficient functions
\begin{eqnarray}
f_1\zeta_1 & = & {}-\frac{2}{5}\, \left( 1 + 55\,{a^2} - 55\,{a^3} -
{a^5} + 30\,{a^2}\,\ln a +30\,{a^3}\,\ln a\right),\nonumber\\
f_1\zeta_2 & = & \frac{1}{36}\left( {a^2} -1 \right) \left( 97 -
648\,a + 1042\,{a^2} - 648\,{a^3} + 97\,{a^4} \right) + \frac{5a^2}{3}
\left( 9 - 16\,a + 9\,{a^2} \right) \,\ln a,\nonumber\\
f_1\zeta_3 & = & {}-\frac{1}{105}\left( 1 - a \right) \,\left( 1349 -
8731\,a - 3817\,{a^2} + 5598\,{a^3} - 3817\,{a^4} - 8731\,{a^5} +
1349\,{a^6} \right)\nonumber\\
& & {}+40\,{a^2}\left( 1 + a \right) \,\left( 3 - 4\,a + 3\,{a^2} \right) \,
   \ln a,\nonumber\\
f_1\kappa_1 & = & {}-\frac{1}{50}\left( 1 - a \right) \,\left( -41 +
284\,a + 3834\,{a^2} + 284\,{a^3} - 41\,{a^4} \right)\nonumber\\
& & {} + \frac{2{a^2}}{5}\, \ln a\left( -80 - 135\,a - {a^3} +
30\,a\,\ln a \right).
\end{eqnarray}
The coefficients $\kappa_2,\kappa_3$ were calculated numerically and are given
in Table~\ref{tab:large} for several values of $a$.

\subsection{The decay $b\to u e \bar\nu$}

In the limit $a\to 0$, the integrals in Eqs.\
(\ref{eq:derAnfang})--(\ref{eq:vomEnde}) can be done analytically, yielding:
\begin{eqnarray}
\lefteqn{D^{virt} = \frac{1}{432}\,\left( 144y^3+180y^2-2540y-513\right) -
\frac{\pi^2}{36}\,\left( y^4-18y^2+16y+6\right)}\nonumber\\
& & {} + \frac{1}{24\sqrt{1-4/y}}\,
\left(4y^3-11y^2-62y+132\right)\,\ln\,\frac{1+\sqrt{1-4/y}}{1-\sqrt{1-4/y}}
\nonumber\\
& & {}-\frac{1}{6}\left(y^4-18y^2+16y+6\right)\!\left\{\ln\,\frac{y}{2}
+ \ln \left( 1+\sqrt{1-\frac{4}{y}}\right)\right\}\!\left\{\ln\,\frac{y}{2}
+\ln \left( 1-\sqrt{1-\frac{4}{y}}\right)\right\}\nonumber\\
& & {}+ \frac{y^2\left(y^2+2y-12\right)}{6}\,\sqrt{1-\frac{4}{y}}\,
\left\{ {\rm L}_2\left[ \frac{1}{2}
\left( 1+\sqrt{1-\frac{4}{y}}\right)\right] - {\rm L}_2\left[\frac{1}{2}
\left( 1-\sqrt{1-\frac{4}{y}}\right)\right]\right\}\nonumber\\
& & {}+ \frac{1}{72}\,\left(12y^3+45y^2-172y-120\right)\,\ln y,\\
\lefteqn{D^{brems} = {}-\frac{1-y}{432}\,(259y^3+43y^2+133y-1863) -
\frac{\ln y}{72}\, (36y^3+477y^2-172y-120)}\nonumber\\
& & {} + \frac{y^4}{12}\,\ln^2 y + \frac{1}{72}\,
(y^4-18y^2+16y+6)\,\left(3\ln^2 y-2\pi^2-12\, {\rm arctan}^2\,
\sqrt{\frac{4}{y}-1}\right.\nonumber\\
& & \left. {} + 24\, {\rm arctan}\,\sqrt{\frac{4}{y}-1} \,\times\,{\rm arctan}
\left( \frac{y}{2-y}\,\sqrt{\frac{4}{y}-1} \right) +
12 \left\{ {\rm L}_2\left[\frac{1}{2}\left(2-y+i y
\sqrt{\frac{4}{y}-1}\right)\right]\right.\right.\nonumber\\
& & \left.\left.{} + {\rm L}_2\left[\frac{1}{2}\left(2-y-i y
\sqrt{\frac{4}{y}-1}\right)\right] \right\}\right) + \frac{\ln y}{6}\,y^2
(y^2+2y-12)\,\sqrt{\frac{4}{y}-1}\nonumber\\
& & {}\times \left\{2\, {\rm arctan}\left( \frac{y}{2-y}\,\sqrt{\frac{4}{y}-1}
\right)+ 2\, {\rm arctan}\left( \frac{1-y}{3-y}\,\sqrt{\frac{4}{y}-1} \right)
+ {\rm arctan} \sqrt{\frac{4}{y}-1} - \pi\right\}\nonumber\\
& & {}+ \frac{y}{36}\,(19y^3+26y^2-234y+72)\,\sqrt{\frac{4}{y}-1}\left\{
{\rm arctan} \sqrt{\frac{4}{y}-1}-{\rm arctan}\left( \frac{y}{2-y}\,
\sqrt{\frac{4}{y}-1} \right)\right\}\nonumber\\
& &{} + \frac{1}{36\sqrt{4/y-1}}\,(19y^4-38y^3-299y^2+678y-468)\,
{\rm arctan}\left( \frac{1-y}{3-y}\,\sqrt{\frac{4}{y}-1} \right).
\end{eqnarray}

\section{Radiative corrections to exclusive decays}
\setcounter{equation}{0}

The $\beta_0^n\alpha_s^{n+1}$ corrections to exclusive decays at
zero recoil were considered in Ref.~\cite{Nnew}, using a somewhat
different approach. To convert to our technique, it suffices to observe
that for ultraviolet convergent quantities the invariant distribution function
$\widehat{w}$ introduced in
\cite{Nnew} is related to the discontinuity of the
one-loop radiative correction $r_0$ with finite gluon mass
$\lambda$ (in the normalization of Eq.\ (\ref{eq:3.39})):
\begin{equation}
r_0(\lambda^2) = \frac{1}{4 C_F}
\int\limits_0^\infty \!\! ds\,\frac{1}{s+\lambda^2}\,
s\,\widehat{w}(\tau = s/\mu^2).
\end{equation}
Using $\mu^2=m_c m_b $ and the explicit expressions\footnote{Note that in
Eq.\ (63) of Ref.~\cite{Nnew} $(1+\tau/2)$ must be replaced by $(1+\tau/z)$.}
given in \cite{Nnew}, we find
(with $y=\lambda^2/m_b^2$ and $z=m_c/m_b$):
\begin{eqnarray}
r_0^V(y) & = & \frac{-3y + 2yz - 4z^2 - 3yz^2}{8z^2}
+ \frac{y\,\left( 1 - z \right)^2\,
   \left( 3y + 4yz - 2z^2 + 3yz^2 \right)}{16z^4}\,\ln y
\nonumber\\
& & {}-\frac{3y^2 - 5y^2z - 2yz^2 + 6yz^3 + 6z^4 + 6z^5}{8z^4(1-z)}\,\ln z
\nonumber\\
& & {}- \frac{3y^3 - 5y^3z - 8y^2z^2 + 16y^2z^3 + 4yz^4 +
   4yz^5 - 32z^6}{16yz^4(1-z)\sqrt{1-4z^2/y}}\,\ln\,
\frac{1+\sqrt{1-4z^2/y}}{1-\sqrt{1-4z^2/y}}\nonumber\\
& & {}- \frac{-4y - 16y^2 + 5y^3 + 32z - 4yz + 8y^2z - 3y^3z}{16 y
(1-z)\sqrt{1-4/y}}\,\ln\, \frac{1+\sqrt{1-4/y}}{1-\sqrt{1-4/y}}\,,\\
r_0^A(y) & = &
{}-\frac{9y + 2yz + 24z^2 + 9yz^2}{24z^2}-\frac{9y^2 - 7y^2z - 6yz^2 - 6yz^3
+ 18z^4 + 18z^5}{24(1-z)z^4}\,\ln z\nonumber\\
& & {}+\frac{y\left( 9y + 2yz - 6z^2 + 2yz^2 - 12z^3 +
     2yz^3 - 6z^4 + 9yz^4 \right)}{48z^4}\,\ln y\nonumber\\
& & {}-\frac{9y^3 - 7y^3z - 24y^2z^2 + 8y^2z^3 + 12yz^4 + 44yz^5 -
96z^6}{48yz^4(1-z)\sqrt{1-4z^2/y}}\,
\ln\,\frac{1+\sqrt{1-4z^2/y}}{1-\sqrt{1-4z^2/y}}\nonumber\\
& & {}-\frac{-44y - 8y^2 + 7y^3 + 96z - 12yz + 24y^2z - 9y^3z}{48y
(1-z)\sqrt{1-4/y}}\,\ln\,\frac{1+\sqrt{1-4/y}}{1-\sqrt{1-4/y}}\,.
\end{eqnarray}


\begin{thebibliography}{99}
\bibitem{inclusive}
The idea to extract $|V_{cb}|$ from inclusive decays is rather old. The
understanding that this may yield a high precision is however very recent,
see \cite{OPE}. More references are given below.
\end{thebibliography}
\end{document}